\documentclass[preprint]{aastex}

\newcommand{\R}{\mathbf{r}}

\begin{document}

\title{Analyzing the designs of planet finding missions}

\author{Dmitry Savransky, N. Jeremy Kasdin and Eric Cady}
\affil{Department of Mechanical and Aerospace Engineering\\ Princeton University, Princeton, NJ 08544}

\email{dsavrans@princeton.edu}

\begin{abstract}
We present a framework for the analysis of direct detection planet finding missions using space telescopes.  This framework generates simulations of complete missions, with varying populations of planets, to produce ensembles of mission simulations, which are used to calculate distributions of mission science yields.  We describe the components of a mission simulation, including the complete description of an arbitrary planetary system, the description of a planet finding instrument, and the modeling of a target system observation.  These components are coupled with a decision modeling algorithm, which allows us to automatically generate mission timelines with simple mission rules that lead to an optimized science yield.  Along with the details of our implementation of this algorithm, we discuss validation techniques and possible future refinements.  We apply this analysis technique to four mission concepts whose common element is a 4m diameter telescope aperture: an internal pupil mapping coronagraph with two different inner working angles, an external occulter, and the THEIA XPC multiple distance occulter.  The focus of this study is to determine the ability of each of these designs to achieve one of their most difficult mission goals - the detection and characterization of Earth-like planets in the habitable zone.  We find that all four designs are capable of detecting on the order of 5 Earth-like planets within a 5 year mission, even if we assume that only 1 out of every 10 stars has such a planet.  The designs do differ significantly in their ability to characterize the planets they find.  Along with science yield, we also analyze fuel usage for the two occulter designs, and discuss the strengths and weaknesses of each of the mission concepts.
\end{abstract}


\section{Introduction}
As the number of known extrasolar planets (exoplanets) continues to grow, so does the desire to directly image these bodies.  Despite the increasing number of ways that have been proposed or implemented to indirectly detect these planets' presence, there is general agreement that direct detection will provide a wealth of new information that is unavailable from most indirect detection methods. The task of repeated direct detections of exoplanets, especially those of the size of the Earth, will most likely require a space-borne observatory.  Another satellite along the lines of the Hubble Space Telescope or James Webb Space Telescope is a major undertaking and should only be built if it has a reasonable chance of succeeding in its stated mission.  In terms of mission planning and design, this translates to formulating a set of requirements which are reasonable given our current state of knowledge, and then designing hardware which will meet these requirements.

To achieve the goal of direct exoplanet imaging, multiple teams are currently developing mission concepts for dedicated planet-finding space observatories.  In this paper, we present a framework for objectively evaluating the expected performance of such direct detection instruments via complete mission simulation  (i.e., generating timelines of simulated observations and their outcomes).  From these, we compute distributions of science yield metrics such as the number of planets found and the number of planetary spectral characterizations.  We apply this framework to four instrument designs with a common observatory and compare the performance of each, demonstrating the utility of this modeling method and hopefully aiding in future decisions as to which technology areas need more development and which mission architecture shows the most promise.

Previous direct detection planet-finding mission studies have focused on specific mission architectures and generally only reported expectation values for a small number of science metrics. 
The basic approach in these studies has been to use a numerically evaluated statistical description of various quantities involved in a planetary detection and report results in terms of expected values  of random variables --- producing either an expected number of planets found as in \citet{brown2005}, or a total number of zones of interest probed as in \citet{lindler2007} (this refers to the fraction of space where a planet of interest might be that is observed in the course of the mission).  While these are powerful metrics for mission performance, they do not map simply to specific mission requirements.  Rather than repeating this for each mission concept considered, we have constructed a general algorithm to simulate an entire mission, along with all detections and observatory operations.  By performing such simulations many times over, with varying populations of planets, we produce an ensemble of mission timelines, giving us distributions of mission science yields. While this is significantly more computationally intensive than earlier treatments, it has the advantage of allowing us to answer a variety of questions about proposed mission architectures with a single ensemble of simulations.  While every new metric (such as, for example, the number and precision of orbital fits achieved) would require an independent statistical formulation, we can simply look at the results of a previously evaluated mission ensemble and extract its distribution.  Additionally, this simulation driven method allows us to evaluate not only different instrument designs, but also different mission rules and observing strategies.


We have previously demonstrated a preliminary implementation of this analysis method in \citet{savransky2008}.  Here, we expand and refine these methods and tools. The first half of this paper (\S\ref{sec:constructDRM}) is devoted to a discussion of each of the components of a mission simulation and the details of our current implementation.  The remainder of the paper describes the applications of this tool to real mission concepts and presents the results of these simulations. \S\ref{sec:validation} is devoted to validating the results of the framework, while \S\ref{sec:compareInstruments} compares the performance of four mission concepts, all of which are geared towards the detection of Earth-like planets.

\section{Constructing Mission Simulations} \label{sec:constructDRM}
The automated construction of a mission simulation requires a mechanism for generating randomized planetary systems and a generalized description of an observation of one of these systems at a given time by a given instrument.  This requires a detailed model of the detection instrument and the observing platform, taking into account all of the operational characteristics of each.  A large number of the mission concepts currently being developed for direct detection of exoplanets  can be categorized into one of two main categories, with many variations of particular details.  The first is a space telescope with an internal coronagraph.  Coronagraph designs include various combinations of shaped pupils \citep{vanderbei2004checkerboard,vanderbei2003circularly,kasdin2005}, apodized pupils \citep{jacquinot1964progress,nisenson2001detection}, Lyot stops \citep{kuchner2002coronagraph,soummer2004apodized}, pupil mapping systems \citep{guyon2003,vanderbei2006diffraction}, etc.. The alternative approach is to fly a separate spacecraft (an `occulter' or `starshade') between the target and the telescope to block the starlight. \citep{cash2006detection,vanderbei2007}

In addition to describing the planets and instruments, a mission simulation also requires a method to determine the order of observations.  Since the scheduling of future observations depends on the result of each current observation, if we want to make each mission simulation completely deterministic and repeatable (rather than including a human in the loop), we also require an algorithm for observation scheduling.  In this section, we will consider each of these three aspects of the simulation in turn.

\subsection{Generation of Planetary Systems}\label{sec:genPlanSys}
For the purposes of this study, we will consider only a highly simplified population of planets---`Earth Twins' on habitable zone orbits.  That is, all of the planets considered will have the mass and radius of the Earth as well as the Earth's average albedo and will reside on orbits with semi-major axes between 0.7 and 1.5 AU (scaled by the square root of their parent star's luminosity) with eccentricities of less than 0.35. \citep{kasting1993,brown2005}  We will also only consider systems having at most one planet. We use the conventional parameter $\eta_\oplus$ to indicate the expected number of Earth-like planets per star.  Since we are limiting our study to one planet per system, in this terminology, $\eta_\oplus = 1/3$ represents a universe where one third of all stars will have an Earth-like planet. \citep{beckwith2008}

We consider this simplistic planetary population because many of the space-based planet-finding mission concepts (including all of the ones in \S\ref{sec:results}) have the stated goal of finding terrestrial planets, which usually represents their most difficult mission requirement. Furthermore, this simple population serves as an excellent proof of concept for our general approach while removing the complexity that goes into modeling realistic planet mass-radius-albedo relations and stable multi-planet systems.  Finally, we note that since our simulation framework is constructed so that all of the fixed parameters above can be changed to any desired distributions, all of these things can be included in future studies with relative ease, incorporating much of the theoretical work currently being done on planet formation and observational statistics. \citep{cumming2003, butler2006, fortney2007}

To populate a simulated `universe', we simply assign planets to some subset of our target list.  For $N$ target stars, a randomly selected subset of size $\eta_\oplus \times N$ is given a planet with randomized average orbital elements. A semi-major axis ($a$)  and eccentricity ($e$) are drawn from uniform distribution in the ranges listed above.  The orbits are oriented in space by generating an argument of pericenter, longitude of the ascending node and inclination ($\omega$, $\Omega$, and $I$).  Since we do not yet have any information that would lead us to expect a non-isotropic distribution of orbital orientations, $\omega$ and $\Omega$ are drawn from uniform distributions in $[0, 2\pi)$ and the inclination is drawn from a uniform distribution in $\cos(I)$ (in the range $[0, \pi)$). \citep{brown2004a} The sets of orbital parameters are converted to position and velocity vectors with respect to the system barycenter ($\R_p$, $\dot{\R}_p$)  at epoch $t_0$, which is taken to be the start date of the mission being simulated. \citep{vinti} 

Finally, the mass of each target star is calculated from its absolute V magnitude via the relation
\begin{equation}\label{eq:lmr}
\log(M_s/M_\odot) = 0.002456V^2 - 0.09711V + 0.4365
\end{equation}
where $M_s$ is the star's mass, and $V$ is the visual magnitude.  This relation is used since it conveniently covers the range of visual magnitudes and stellar masses considered here.  However, newer mass-luminosity relations exist, with better accuracy for different mass ranges, and will be incorporated in future implementations. \citep{reid2002}  This empirical relation has been shown to be accurate to within 7\% for mass ranges corresponding to visual magnitudes between 1.45 and 10.25. \citep{henry1993}  We treat the output of equation (\ref{eq:lmr}) as the `estimated' mass and generate a `true' stellar mass as
\begin{equation}
M_{true} = M_{est}(1+ 0.07\nu)
\end{equation}
where $\nu$ is a uniformly distributed random number in the range $[-1,1]$.  The true masses are used for propagating the planets along their orbits, while the estimated masses of stars are needed for our return strategies, as described in \S\ref{sec:decisionModeling}.  We assume an error in our knowledge of the target stars' masses since knowledge of these is important to our return visit strategies, and we cannot expect perfect knowledge of the masses of nearby stars in the foreseeable future.

At the end of this process, we have a universe encoded as an array of $N$ parameter sets
\begin{displaymath}
\{\begin{array}{ccccccccc} M_{true} & M_{est} & \R_p & \dot{\R}_p & \R_s & \dot{\R}_s \end{array} \}
\end{displaymath}
where $\R_s, \dot{\R}_s$ are the position and velocity of the star with respect to the system barycenter.
 
\subsection{Modeling an Observation}\label{sec:modelObs}
An observation of (or `visit' to) a target system can be broken down into two main activities: first, it is necessary to determine whether a planet has been detected; second, in the event of a detection, the planet may be characterized using the specific capabilities of the instrument (here, we will consider only spectral characterization, with orbital characterization obtained as a result of multiple detections of the same planet).  We model the success or failure of both these tasks based on the time required compared with the time available.  An observation will result in one of four possible outcomes: a detection, a null detection, a missed detection or a false alarm.  A null detection signifies that there is no observable planet in the field of view at the time of the observation, whereas a missed detection occurs when an observable planet exists, but not enough time is spent integrating to achieve a detection.  A false alarm indicates that noise or other objects in the field of view are mistaken for a planet. In the case of false alarms, it is assumed that followup spectroscopy will be able to resolve these as null detections (at the cost of additional integration time).\footnote{In statistical terms, a null detection is a true negative, a missed detection is a false negative, a false alarm is a false positive, and a detection is a true positive.}

In our simulations, observations of systems with planets will generate detections, missed detections or null detections, while observations of systems without planets will generate null detections or false alarms.  In reality, false alarms can occur in systems with planets, however since the current simulations assume that any false alarm will be followed immediately by spectroscopy, any observable planet will necessarily be discovered in this process, changing the false alarm to a detection.  In the future, if we wish to simulate scenarios without a method of immediately resolving false alarms, then false alarms will have to be included for systems with planets.

In order to calculate how much time is required to accomplish a detection and spectral characterization, we must describe the capabilities of a direct detection instrument.  The modeling of an arbitrary planetary observation has been previously described; see, for example, \citet{brown2005} or \cite{savransky2008}.  We will outline these results here to standardize the notation used for the remainder of this paper. Briefly, a direct observation of a planet produces two pieces of information.  The first is the apparent separation $s$, which is the magnitude of the star-planet vector projected into the plane of the sky,
\begin{equation}
s = \left\Vert \left[\begin{array}{ccc}1 & 0 & 0 \\ 0 & 1 & 0 \\ 0 & 0 & 0\end{array}\right] \R_{p/s} \right\Vert
\end{equation}
where $\R_{p/s}$ is the star-planet vector (i.e. $\R_{p/s} = \R_p - \R_s$).  The second piece of information is the relative brightness of the planet as compared with the star, which is commonly represented as the difference in magnitude between star and planet ($\Delta$mag).  This value can be found via
\begin{equation}
\Delta\textrm{mag} = -2.5\log\left(p \left(\frac{R}{\Vert \R_{p/s} \Vert}\right)^2 \Phi(\beta)\right)
\end{equation}
where $\Phi$ is the planetary phase function,  and $\beta$ is the phase (star-planet-observer) angle. \citep{sobolev}  This angle is closely approximated as
\begin{equation}
\beta \approx \tan^{-1}\left(\frac{s}{z}\right)
\end{equation}
where $z$ is the remaining component of $\R_{p/s}$ after it is projected onto the plane of the sky.  

Because all of these values can be calculated from the star-planet vector, and do not depend on any specific characteristics of the target star, it is possible to calculate the probability of detecting a planet, given that one exists, at a specific apparent separation and $\Delta$mag for any distribution of orbital and planetary parameters.  The cumulative distribution function of this probability distribution is known as the single visit completeness and is described in great detail in \citet{brown2005}. By mapping instrument specifications to limiting values for $s$ and $\Delta$mag, we can use the single visit completeness to report the probability of detecting a planet with a given instrument at one observational epoch, assuming the planet exists in the system being observed.  This mapping is done by assuming that there exists a minimum angular separation between the star and planet (known as inner working angle or IWA) at which the planet can be detected. Thus, the minimum observable apparent separation is equal to the projected IWA (IWA$ \times d$ where $d$ is the distance to the target star)\footnote{If $d$ is in parsecs and the IWA in arcseconds, this gives a value for the projected IWA in AU.}. The limiting $\Delta$mag is the brightness difference between a star and planet beyond which the planet is always unobservable, regardless of their absolute brightness.  Oftentimes, this value is equated with the designed contrast of an instrument (i.e., the ratio of the core to halo of a coronagraph's point spread function). \citet{brown2005} defines this limit as the point where systematic errors produce unresolvable confusion between the planet signal and the background.  

It is important to note that single visit completeness (referred to simply as `completeness' for the remainder of this discussion), does not easily map to the probability of detection over multiple visits.  If each observation was fully independent, with constant detection probability ($p_c$) equal to the single visit completeness, then the probability of $k$ successful detections in $n$ visits would be  given simply by the binomial theorem as
\begin{equation}
P_n(k) = \left(\begin{array}{c} n \\ k \end{array} \right) p_c^k (1-p_c)^{n-k} \, .
\end{equation}
The probability of any (non-zero) number of detections in $n$ visits would then be
\begin{equation} \label{eq:prob}
P_n(k>0) = 1 - (1-p_c)^n\, .
\end{equation}
In reality the probability of detecting a planet on subsequent visits is conditionally dependent on the planet's orbit and the times between observations, but this equation gives a good upper bound for the probability of detecting a planet (given that one exists) over multiple visits.  

The exception to this bound would be the case of an optimal observing strategy (i.e., one where we only see previously unobservable portions of the habitable zone on each subsequent visit) on a fully observable system (i.e., one where all parts of the habitable zone are observable at some point).  In this scenario, a new portion of the habitable zone equal to the fraction represented by $p$ is observed at each visit and the schedule is such that each portion is observed in sequence so no planets could be bypassed (the equivalent of a continuous observation).  If such an observing schedule could be achieved, the probability of detecting a planet after $1/p$ visits would be unity.  For the majority of targets, however, it is quite difficult to follow the strict timing dictated by this strategy.  Because target stars may only be observed for specific intervals of time at given times of the year (during their `observing season', which is determined by an observatory's keepout zone, discussed in \S\ref{sec:decisionModeling}), it is often impossible to follow the optimal observing schedule in the limited period defined by the mission lifetime.  Because of this, we often fail to detect existing planets even when the target system has been visited enough times to yield a high probability of detection via equation (\ref{eq:prob}).  Simulation shows that this equation holds as an upper bound for systems with permanently unobservable sections of the habitable zone (where the habitable zone is partially inside the projected IWA).

 \bigskip
 
Having modeled the data gathered during an observation, we can now calculate require integration times, thereby deciding whether a given observation will lead to a detection.  Following \citet{kasdin2006}, we assume that the instrument produces images with sufficient sampling so that matched filtering (or some other probabilistic detection algorithm) can be applied, which significantly reduces the integration time required to decide whether a planet is present in the field of view.  We treat the photons received at pixel $j$ of the detector array as a random variable of the form
\begin{equation}
z_j = C_p \bar{P}_j + C_b  + \nu
\end{equation}
where $C_p$ and $C_b$ are the mean photon count at the pixel centered on the point spread function (PSF) and mean background photon count at all pixels, respectively, $\bar{P}$ is the normalized, non-dimensionalized PSF, and $\nu$  is the photon and readout noise.  This allows us to construct a signal to noise metric as the random  variable $\hat{C}_p/\sigma_b$ where $\hat{C}_p$ is the linear, unbiased estimate of the planet signal, and $\sigma_b$ is the variance of this estimate for pixels without a planet.  The assumption made in this metric is that the background contribution is known, or can be estimated during the integration.

Simplifying the statistics by assuming the photon arrival rate can be approximated as a Gaussian distribution,  and noting that both $C_p$ and $C_b$ are linear in integration time ($t$), we can write an expression for $t$ based on a desired false alarm probability (FAP) and missed detection probability (MDP).  \citet{kasdin2006} derive this expression as
\begin{equation}\label{eq:detIntTime}
t = \frac{1}{b} \frac{(K - \gamma \sqrt{1 + \tilde{Q} \Xi/\Psi})^2}{\tilde{Q}T_A \Psi}
\end{equation}
where $\tilde{Q}$ is the ratio of $C_p/C_b$ ($\triangleq Q$) multiplied by the sum of $\bar{P}$,  $\Xi = \sum_j \bar{P}^3_j/(\sum_j \bar{P}_j)^3$, and $\Psi = \sum_j \bar{P}^2_j/(\sum_j \bar{P}_j)^2$ (this parameter is also known as the `sharpness').  $K$ is a threshold value determined by the required FAP, such that $\Phi(K) = 1 - \textrm{FAP}$, where $\Phi(\cdot)$ represents the Gaussian distribution function.  Similarly, $\gamma$ is the threshold determined by the required MDP, such that $\Phi(\gamma) = \textrm{MDP}$.  Finally, $T_A$ and $b$ are derived from properties of the optical system and target star.  $T_A$ is the `Airy Throughput': the starlight supression system throughput ($T$) multiplied by the non-dimensionalized pixel area and the sum of the normalized PSF \citep{vanderbei2003circularly} and $b$ is defined as
\begin{equation}
b = \textrm{QE} \, \upsilon I_p(\lambda_r) \Delta \lambda A
\end{equation}
where $\textrm{QE}$ is the detector's quantum efficiency, $\upsilon$ is the attenuation of optical elements up to the exit pupil, and $A$ is the entrance aperture area.  The term $I_p(\lambda_r) \Delta \lambda$ represents the total irradiance of the planet in the detection band (centered at $\lambda_r$ with bandwidth $\Delta \lambda$). $I_p(\lambda_r)$ is the average irradiance in this band, and is calculated as
\begin{equation}
I_p(\lambda_r) = \mathcal{F}_010^{-(V_s + \Delta\textrm{mag})/2.5}
\end{equation}
where $V_s$ is the target star's apparent magnitude in the detection band and $\mathcal{F}_0$ is the band specific flux for a zero magnitude star.  For detections in the V-band (which all systems under consideration here assume), this value is approximately 9500 photons/cm$^2$/nm/s. \citep{colina1996}

Equation (\ref{eq:detIntTime}) allows us to calculate how much time we need to spend integrating before we can assume, with high confidence, that no planet exists (or is currently observable by our instrument).  In our simulation framework, we calculate the required integration time for each target star assuming a planet observed at the IWA, which we define as the 50\% peak throughput point for instruments where the throughput is a function of angular separation\footnote{The throughput discussed here relates only to the starlight suppression system and ignores any downstream optics.} \citep{guyon2006}, and the instrument's limiting $\Delta$mag.  For missed and null detections, this time value is used as the actual integration time (in a real mission, this would be the time spent integrating before moving on to the next target).  For detections, the integration time is recalculated based on the simulated planet's orbital position, using the actual separation (and throughput) and $\Delta$mag of the planet at the time of the observation (in a real mission, this would be the point when a detection could be confirmed via a bayesian technique and integration would either be stopped, or switched to spectral characterization).  False alarms (when no observable planet exists) and missed detections (when there is an observable planet) are generated at the rate determined by the FAP and MDP values used in calculating the integration time in equation (\ref{eq:detIntTime}).  We set these to 1\% divided by the expected number of observations for the whole mission, and 0.1\%, respectively.  These values yielded zero missed detections and at most one false alarm per mision simulation for all of the results in \S\ref{sec:results}.

\begin{figure}[ht]
   \plotone{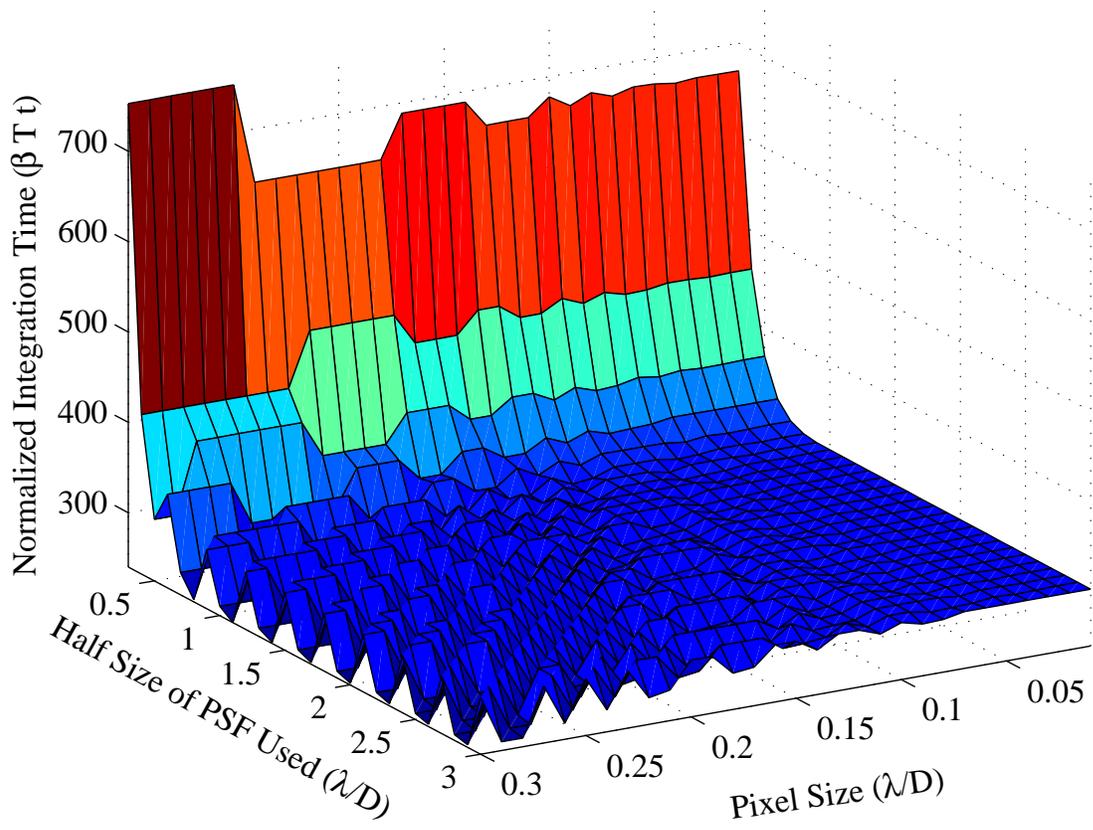}
 \caption[]{ \label{fig:intTimevPSF} Normalized integration time as a function of half size PSF used and pixel area for an open circular aperture.}
 \end{figure}

As in \citep{kasdin2006}, we can consider what effects the pixel size and extent of the PSF used have on the integration time calculation.  Figure \ref{fig:intTimevPSF} shows the integration time for constant $b$ and $T$ as a function of size of the PSF used and the pixel area.  We see that there exist critical values for each of these beyond which the integration time is constant.  In fact, as long as at least 1.5 $\lambda/D$ of the PSF are sampled, changes in integration time due to different pixel sizes are relatively minor.  The same is found when considering non-Airy PSFs.

To calculate the value of $Q$, which was previously defined as the ratio of $C_p$ to $C_b$, we first define the value
\begin{equation}
c_1 \triangleq  \textrm{QE} \, \upsilon \Delta\lambda A T t \,.
\end{equation}
Then
\begin{equation}
C_p = \mathcal{F}_0 10^{-(V_s + \Delta\textrm{mag})/2.5} c_1
\end{equation}
and
\begin{equation}
C_b = \mathcal{F}_0 10^{-V_s/2.5} C c_1+ \mathcal{F}_0 10^{-\Omega_{zodi}/2.5} \left(f(\bar{\beta}) +\mu f(I)2.5^{4.78 - M_V} \right) \Delta \alpha c_1 + \textrm{DR} t + \frac{\sigma_r^2}{t_r}t
\end{equation}
where $C$ is the instrument's designed contrast,  $\Omega_{zodi}$ is the detection band intensity of local zodiacal light in magnitudes per square arcsecond at the ecliptic pole (nominally,  23.34 for V band), $M_V$ is the absolute magnitude of the target star, $\mu$ is the brightness of the extrasolar zodiacal light (exo-zodi) in units of zodi, $\sigma_r$ is the detector readnoise, $t_r$ is the exposure time per readout, DR is the detector dark current rate per pixel, and $\Delta\alpha$ is the pixel area on the sky in square arcseconds.  The pixel area can be written as
\begin{equation}
\Delta \alpha =\Omega\left(\frac{180*3600}{\pi}\right)^2
\end{equation}
where $\Omega$ is the solid angle of a detector pixel in steradians and equals $(\lambda/2D)^2$ for critically sampled systems with circular apertures. \citep{brown2005}  The function $f$ is the empirically derived variation of zodiacal light with viewing angle given by
\begin{equation}
f(\theta) = 2.44 - 0.0403 \theta + 0.000269 \theta^2
\end{equation}
with $\theta$ in degrees, in the range [0,90] (the relation is mirrored for $\theta \in [90,180]$).  The function is applied to $\bar{\beta}$ - the absolute value of the ecliptic latitude of the target star, and $I$ - the target star system's inclination. (D. Lindler,  Personal Communication, 2008) 

 It should be noted that a major assumption being made here is that the exo-zodi is uniform and of constant magnitude in the region being observed.  For the results in this paper, we assume a constant exo-zodi level higher than that of the solar system to prevent overly optimistic results, but an important future consideration is how the possibility of non-uniform exo-zodi will increase the required integration times for detection.  Putting all this together, we evaluate $Q$ as
\begin{equation}
Q = \left[ 10^{\Delta\textrm{mag}/2.5} C + 10^{(V_s +\Delta\textrm{mag}-\Omega_{zodi})/2.5} Z \Delta\alpha + \frac{ 10^{(V_s + \Delta\textrm{mag})/2.5} }{\mathcal{F}_0 \textrm{QE} \, \upsilon \Delta\lambda A T}\left(\textrm{DR} + \frac{\sigma_r^2}{t_r}\right) \right]^{-1}
\end{equation}
 where $Z = \left(f(\bar{\beta}) +\mu f(I)2.5^{4.78 - M_V} \right)$.
 
In the event of a detection, we calculate the time required for spectral characterization, assuming that the detection integration provides us the magnitude of all background sources.  Following \citep{lindler2007}, we will represent spectral characterization requirements as minimum signal to noise (S/N) values, using the metric
\begin{equation}
\textrm{S/N} = \frac{C_p}{\sqrt{V}}
\end{equation}
where $V$ is the total variance due to all noise sources.  This value can be written as a function of the square root of the integration time as
\begin{equation}
\textrm{S/N} = \frac{\tilde{C_p} \sqrt{t}}{N_{pix}\left( \left(\frac{\sigma_r^2}{t_r} + \textrm{DR} \right) \left(1 + \frac{1}{N_{dark}}\right) + \mathcal{F}_0 10^{-\Omega_{zodi}/2.5} Z \Delta\alpha \right) + \tilde{C}_p + \tilde{C}_s }
\end{equation}
where $\tilde{C}_p$ and $\tilde{C}_s$ are the planet and star counts per unit time, $N_{dark}$ is the number of dark frames used, and $N_{pix}$ is the number of pixels in the detection box.  The specific S/N value used depends on the spectral feature of interest.  For terrestrial planets, we are especially interested in the possible discovery of biomarkers such as oxygen or ozone.  Several of these features are identified in \citet{heap2007}, along with their S/N requirements; for example, a S/N of 11 for a resolving power of $R = 70$ will produce detections of oxygen at Earth atmospheric levels (21\% ) via the feature at 760nm with a confidence level of over 99.9999\%.  For 20\% Earth O$_2$ abundance, the same signal to noise will still give confidence levels of over 99\%. \citep{desmarais2002}.  This is the target S/N level used for spectra in \S\ref{sec:results}.
 
\subsection{Decision Modeling} \label{sec:decisionModeling}
Having expressed all of the quantities involved in an observation in terms of the astronomical values generated in \S\ref{sec:genPlanSys}, and instrument characteristics described in \S\ref{sec:modelObs}, we now need an algorithm for selecting the order in which targets will be observed.  Our goal is to optimize the science yield of a mission by simultaneously maximizing the number of unique planets discovered, the number of spectral characterizations of discovered planets, the number of planets observed enough times to produce orbital fits, and the portion of the target list observed at least once.  This last goal is included since there is scientific merit in any observation, whether or not a planet is detected (in the form of disk science and concurrent observations by other instruments on the observatory), and this provides a simple way of controlling the decision algorithm's natural bias towards higher completeness stars.

 Previous implementations of automated planning have modeled the order of observations as a traveling salesman problem (TSP) with repeated visits---a useful model as there exists a large body of work on methods for solving it. \citep{kolemen2007} Unfortunately, it is difficult to create a realistic mission simulation with this approach.  Unlike the classical TSP, where path lengths between locations are fixed and the group of locations remains invariant, here, the cost of transferring between a pair of target stars will depend on the location of the spacecraft, and the subset of available targets will itself change in time, due to the keepout zones associated with various portions of the system.  Even worse, the amount of time the spacecraft must spend on a given target will depend on whether a detection occurs, which cannot be known \emph{a priori}.  For these reasons, even the time-dependent TSP description of the system is inadequate.

Instead, as with a TSP, we represent the sequence of visits as a directed graph with variable length edges.  The graph is encoded as an $N \times N$ matrix (for $N$ targets), where element $ij$ represents the `cost' associated with switching from target $i$ to target $j$ (which we shall refer to as `transiting' between targets).  We can use a variety of graph search techniques to determine the optimum (least cost) path, by assuming that this matrix will remain constant for $k$ transits (where $k$ is determined by the specifics of the system being simulated), and thus pick the next target as the node generating the least cost path.  After a target is observed, however, the entire matrix is re-generated based on the current location of the spacecraft, and the process is repeated, thereby capturing the dynamic nature of the problem.  We can also take the approach of simply always going to the next best available target, by setting $k = 1$.  In cases where the matrix is relatively stable over multiple steps (i.e., the time spent on each target is small), the re-evaluation of the costs does not produce significant changes in the planned path.  However, in those cases where one observation takes a long time (i.e., when spectral characterization is required), the costs will change significantly, so it is very important to update the matrix to select the next best target.

The function determining the cost of each transit is a weighted linear combination of multiple factors,
\begin{equation}\label{eq:costfunc}
A_{ij}  = \left[ \begin{array}{c}  
a_1 \frac{\cos^{-1}\left(u_i \cdot u_j\right)}{2\pi}B_{inst} + a_2 \textrm{comp}_j - a_3 e^{t_c-t_f} B_{unvisited} + \\ 
\medskip
 a_4 B_{visited}(1-B_{revisit}) -a_5 B_{revisit} \left(\frac{N_j}{N_{req}} \right)(N_j < N_{req}) - a_6 \frac{\tau_j}{\textrm{vis}_j}
\end{array} \right] (1-B_{keepout}) \, .
\end{equation}
The first term represents the cost associated with retargeting, with $B_{inst}$ set equal to 0 for coronagraphs and 1 for occulters (since the amount of fuel and time used in retargeting should be approximately constant for an internal coronagraph for any pair of targets).  While this term is not the actual cost associated with each specific retargeting (which depends on the position of the spacecraft and its orbit), it serves as a good heuristic function.  Because the orbits assumed for the telescope and occulter are such that active control is required to retarget in a reasonable amount of time, and the contribution of the orbital dynamics to the time and fuel costs of retargeting are comparatively small, in most cases this represents an admissible heuristic, and, as shown in \S\ref{sec:validation}, produces good results. \citep{kolemen2007} In calculating the actual transit time and fuel used, the full orbital dynamics and spacecraft masses are taken into account, and the masses are updated after each transit to reflect fuel use.  Briefly, following \citet{kolemen2007}, transits are modeled either as impulsive thrusts or continuous point to point trajectories.  In each case, the aim is to find an occulter trajectory such that the telescope-occulter separation is fixed at the end, with inertial velocities matched. In the impulsive case, this is achieved with two large maneuvers at the beginning and end of the trajectory which are taken to change the occulter's velocity while keeping its position nearly constant.  This becomes a boundary value problem with the 3 dimensional velocity vectors at the trajectory endpoints as the unknowns, which can be solved via collocation.  In the case where longer thrust times are required (such as with electric propulsion), the strategy is to minimize the control effort given the differential equations of motion and fixed endpoints (a fixed transfer time is assumed), which results in a 12 dimensional boundary value problem.  For more details, please see \citet{kolemen2007} and \citet{stengel1994optimal}.  Fuel use is calculated by assuming that the spacecraft mass is nearly constant during any impulsive thrust maneuver and taking the product of the burn time with the propulsion system mass flow rate, which is determined by the system specific $I_{sp}$ and thrust force.

The second term serves as a heuristic for the probability of a detection at the next target, with  $\textrm{comp}_j$ equal to one minus the completeness of the $j$th target.  The third term is included to promote visits to as many unique targets as possible.  Here,  $t_c$ is the current time, $t_f$ is the mission lifetime, and $B_{unvisited}$ is a boolean equal to 1 if target $j$ has not been visited.  The fourth term creates a preference for targets that are scheduled for a revisit, with boolean $B_{visited}$  set to 1 if target $j$ has been visited, and boolean $B_{revisit}$ set to 1 if target $j$ is currently scheduled for a revisit (`currently', in this case, means $\pm$ the average re-targeting time).  The fifth term is included to improve the odds of good orbital fits.  $N_j$ is the number of previous detections of a planet at target $j$, and $N_{req}$ is the number of detections required for an orbital fit.  This term has the effect of biasing target selection towards those stars with more prior detections as they near the minimum required number of detections, when they are scheduled for revisits.  The inequality expression represents a boolean that eliminates the effect of this term once the required number of detections has been achieved.  For this study, we have taken $N_{req}$ to be four, but depending on the system and desired accuracy of derived orbital elements, many more detections may be required.  Since the method used to schedule future observations (see below) depends on an estimate of the semi-major axis, which is updated with each subsequent observation, we have found that four detections are generally enough to constrain the semi-major axes of planets in the limited population considered here to below 10\% error (in some cases to within 1\%) as well as constraining the possible ranges of the other orbital parameters, giving us the ability to say with high confidence whether the planet is in the habitable zone.  Nevertheless, previous work has shown that orbital fitting accurate enough to predict the future positions of planets is very difficult, and may require many more observations, especially when considering more diverse populations of planets.  For more details, see \citet{pravdo2007observation} and \citet{brown2007minimumTPF}. 

The final term inside the brackets was included after study of strategies employed in the manual scheduling of planet finding missions.  The most valuable of these is identifying the stars closest to `leaving' a keepout zone, and thus providing the maximum amount of possible integration time for the instrument.  This has the additional benefit of creating a schedule in which an occulter does not need to move from target to target at full thrust (since slew times are determined by how much time is left until the star is viewable), thus saving large amounts of fuel.  This behavior is modeled by the sixth term, where $\tau_j$ is the amount of time before star $j$ enters a viewable zone (this value is negative if the star is already outside of the keepout zone) and $\textrm{vis}_j$ is the total amount of time star $j$ is continuously viewable. This term has the effect of making stars about to enter a viewable portion of the sky more likely to be selected, while penalizing those that are about to enter a keepout zone. 

The factors $a_i$ are selected to weight the relative importance of the various terms.  For the results presented in the next section, these terms were chosen manually, after experimenting with varying values.  Obviously, these represent a great opportunity for optimization and possibly auto-tuning during the course of a single simulation.  However, even with only six parameters, the search space of possible values is quite large, making an optimization remarkably computationally intensive; we reserve this for a future study.  Finally, $B_{keepout}$ is a boolean representing whether a given star is currently in the keepout zone.  For any given telescope, the sun cannot be within some number of degrees of the line of sight (spacecraft - target vector) or light from the sun would enter the telescope aperture (for this study, we take this value to be 45$^\circ$ for all instruments considered).  For coronagraphs, this is the only factor determining whether a star is in the keepout zone.  For occulters, an additional keepout zone is required to prevent reflection of sunlight from the occulter into the telescope.  This corresponds to the 180$^\circ$ region below the line orthogonal to the line of sight (i.e., in this case, the sun can only be in two 45$^\circ$ regions to either side of the telescope - occulter - target vector).  For an occulter system, stars are only considered to be in the keepout zone if they will be unobservable when the occulter completes its transition slew.  That is, if the amount of time until they become viewable would result in a transit slew using less than some selected factor of the slew time for thrusting (in our study, we used 50\%), or if they could not be reached at full thrust before entering the keepout zone.  

For the instruments considered in this paper, those with an occulter will have transfer times between targets on the order of weeks, so that $k$ must be quite small---just two or three steps.  This makes it easy to compute all possible path costs and take the one representing the true minimum.  Self-contained coronagraphs, on the other hand, can acquire a new target in as little as a single day, so that it would be possible to use a much larger $k$, which would require switching to a different selection method (such as an iterative deepening search).  However, simulation shows that due to the relatively small target pools available, in most cases, just using the first 3 to 5 steps produces the same result as a more exhaustive search.  Thus, to minimize computing time, our implementation uses a fixed $k$ of 3 for these instruments, while allowing for the possibility of making this parameter variable in the future.

Revisit scheduling depends on whether a detection occurs.  As shown in \citet{savransky2007}, if a detection occurs, the most likely estimate for the planet's orbital period is obtained by assuming that the observed apparent separation is equal to the planet's semi-major axis and then calculating the star's mass from its apparent magnitude, as in equation (\ref{eq:lmr}).  The best time to schedule a return (i.e., the most likely to result in a repeat detection) is either a full or half estimated orbital period.  We choose to use half of the estimated orbital period in order to get repeat detections on separated positions on the orbit, to improve the orbital fit.  If no detection occurs, we rely on the mean orbital period of the planet population of interest, and schedule the return visit for 3/4 of this mean period.  Every subsequent detection is used to improve our knowledge of the planet's orbital semi-major axis.  Note that the apparent separation can be less than the semi-major axis due to either orientation of the orbit with respect to the line of site or the orbit's eccentricity.  It can be greater than the semi-major axis due only to eccentricity.  We use the probability density function of the parameter defined as the ratio of $s/a$ to weight each recorded $s$ in order to update our estimate of $a$. \citep{savransky2007}  The list of scheduled revisits is used to set the $B_{revisit}$ term in equation (\ref{eq:costfunc}).	

It is important to point out that the return visit strategy may be significantly affected by the detection of multiple planets during the same visit.  If multiple planets are seen at wide angular separations during one observation, assuming that they have small mutual inclinations, it might be safe to assume that the exosystem was being viewed nearly pole-on, and thus would have already captured about as much of the habitable zone as possible, thereby negating the need for future visits if you were only interested in detection and spectral characterization (of course, multiple visits would still be needed for orbital characterizations).  Attempting to estimate the frequency at which multiple detections will occur during single observations is fairly difficult, since it depends completely on the makeup of exosystems.  Our currently small sample of multiple planet system contains a large fraction of `hot jupiters' due to the biases of the detection methods used.  These planets, due to their proximity to their primary stars, would not be detectable with the instruments studied here, and thus do not constitute a good test.  Instead, we can take our own solar system, and simulate randomly timed observations with the plane of the ecliptic randomly oriented with respect to the line of sight, and the system barycenter located a random distance from the observer.   Doing so for distances between 10 and 30 parsecs, with observations randomly placed in a 5 year window, and using a 75mas IWA instrument with a limiting $\Delta$mag of 26, produces slightly over 30\% of observations in which more than one planet could be seen, with only 4\% in which more than two are detectable.  Furthermore, because Jupiter and Saturn are so highly detectable, the system orientation was close to pole-on (less than 10 degrees rotation) in only 20\% of these cases.  Thus, while it is probable that multiple detections will occur within single observations of systems structured somewhat like our own, this does not guarantee that we will be able to discount systems after a  single visit.  Even in cases where we will have good constraints on the exosystem inclination (i.e., when at least one of the planets detected is clearly in the inner system), one could argue that repeat observations as scheduled by the algorithm used here can be useful.  First, they will likely be required for confirmation---re-detecting a planet when it has moved noticeably on its orbit makes a much stronger case that the original detection was not of a background object or structure in the zodiacal dust. Second, while our solar system has very small mutual inclinations, we cannot discount the possibility that there exist stable exosystems with high mutual inclinations \citep{libert2007exoplanetary}.  Future implementations of this framework will incorporate multi-planet systems, but we believe the results of the single planet system trial are still useful in evaluating and comparing the performance of planet-finding instruments.

\begin{figure}[ht]
   \plotone{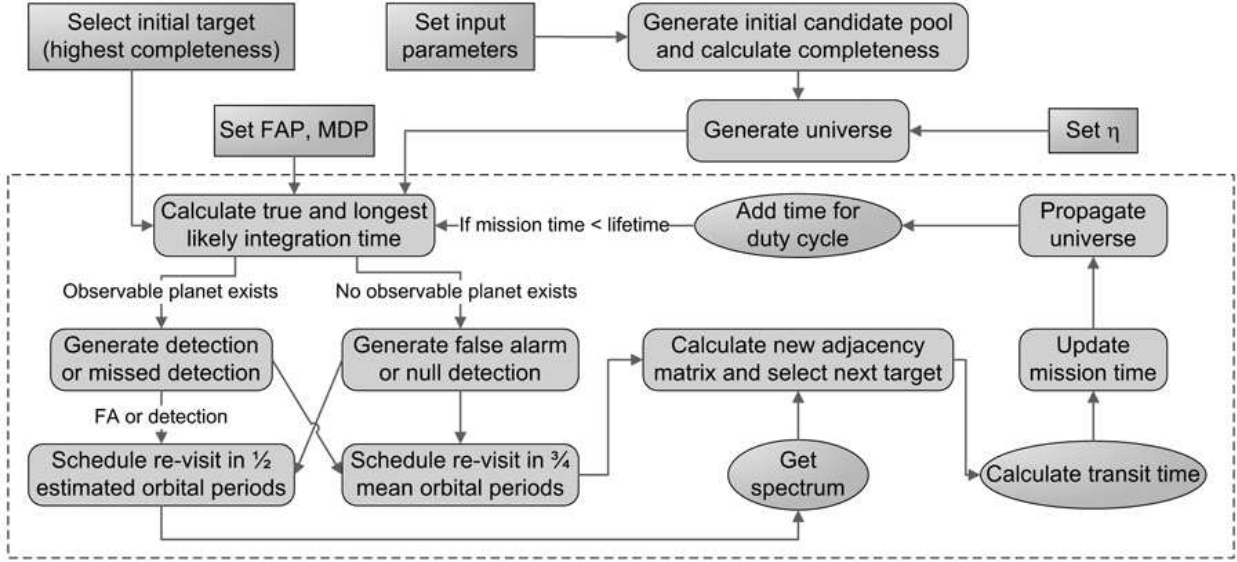}
 \caption[]{ \label{fig:flowchart} Flowchart of simulation framework.  Ellipses represent optional steps determined by the specific instrument or rule set.}
 \end{figure}

\begin{figure}[ht]
   \plotone{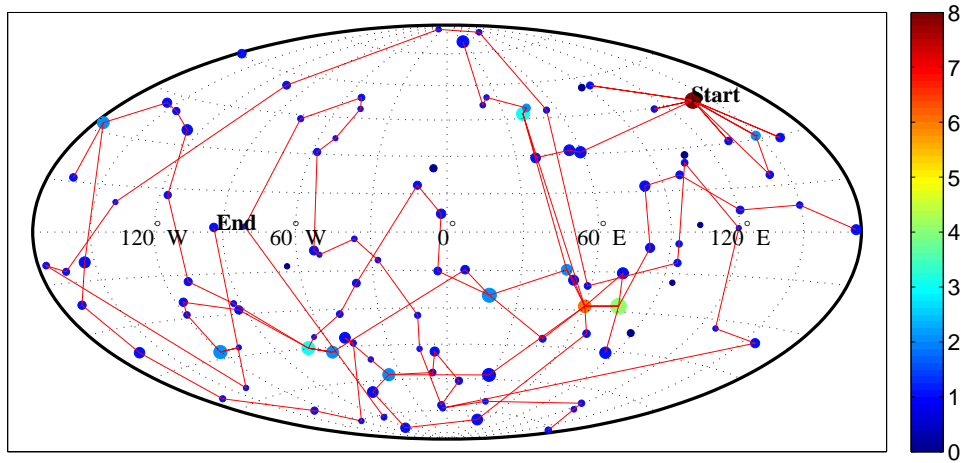}
 \caption[]{ \label{fig:theia_DRM} Visualization of an automatically generated sequence of observations. Points represent target stars, with size correlated with single visit completeness (i.e., the larger the point, the higher its single visit completeness). The color scale represents the number of observations made of a system in this simulation.  Note that some systems get many visits not because of their high completeness, but because they are well located on the sky.}
 \end{figure}

Figure \ref{fig:flowchart} shows a flowchart for our implementation of the simulation framework.  Propagation of the simulated `universe' is achieved via an N-body integration of the state vectors composed of the planet and star position and velocity vectors for each system.  In our implementation, this is done using a Runge-Kutta-Nystr\"{o}m symplectic variable time-step scheme.  To check whether the path generated is reasonable, and to allow for easier visualization of a whole mission timeline,  we can plot the sequence of observations on a map of the sky.  A sample of this is shown in Figure \ref{fig:theia_DRM}.

\subsection{Target Lists} \label{sec:targlists}
Finally, we consider how to select the pool of target stars from which the simulation determines the visit order.  Because many of the specific results in \S\ref{sec:results} are driven by the fact that we have a limited population of stars, it is very important to identify exactly which stars will be included in any given mission simulation.  We want to start with a list of all of the stars about which a planet from the planetary population of interest, on an orbit of interest, could, at some point in its orbit, be observable by our instrument.  Since we are currently interested in Earth-twins in the habitable zone, we leave only those stars where a randomly oriented habitable zone orbit leaves the planet outside the instrument's projected IWA for a time longer than the required detection integration time for that star (as defined in \S\ref{sec:modelObs}).   The procedure for generating these lists involves finding the smallest possible $\Delta$mag corresponding to a value of $s$ greater than the projected IWA and is developed in \citet{savransky2008}. For the simulations in \S\ref{sec:results}, we start with the 1612 non-binary  stars within 30 pc of our solar system and apply the procedure above to the subset of main sequence stars to get 372 possible target stars.  Of these, 144 have completeness values of less than 0.01, and 260 have completeness values of less than 0.1.

The biggest problem in selecting the target list comes from one of the inherent trade-offs in the stated science goals.  The desire to visit as many unique targets as possible means that the overall probability of detecting planets is decreased by including many low completeness stars.  On the other hand,  only visiting a small number of stars automatically limits the number of unique detections you can make, and in low $\eta_\oplus$ universes, having a large target pool may be the only way to detect any planets at all.  An additional problem lies in trying to filter target stars by the amount of integration time required.  Although we can calculate the integration time for any assumed $\Delta$mag value, in reality, planets from our population of interest can vary by several magnitudes at the time of observation.  For example, a planet observed at one point on its orbit at a $\Delta$mag of 26 will require 200 days for spectral characterization.  The same planet, at a different orbital position can have a $\Delta$mag of 25, which would require only 30 days of integration.  In both cases, the time required to determine whether a detection has occurred is less than 2 days.  Because of the great disparity in detection and characterization times, and since high ecliptic latitude stars have very long observing seasons, the basic mission simulation logic can sometimes lead to extremely long integration times which adversely affect overall mission performance.

Because we do not assume \emph{a priori} knowledge of $\eta_\oplus$ or what the brightness of planets will be at the moment of detection, it is quite difficult to address these issues using only the transition cost function.  Instead, we introduce two additional parameters - a minimum target system completeness value, and a maximum single integration time.  The minimum completeness gives us a systematic way to limit the size of target lists, mapping directly to the unique targets vs. total detections tradeoff. The maximum integration time allows us to include all potential targets, but not allow any one integration to significantly affect the mission outcome.  This value is used in two different ways: first, the target list is filtered so that no detection time for the worst case (highest zodi level and limiting $\Delta$mag) exceeds the minimum integration time.  Second, in visits where a detection occurs, but the spectral characterization would take more than the maximum integration time, the spectral characterization is not attempted. 

The value for each of these is found by performing a line search over multiple simulation with these values smoothly varied to optimize our target metrics.  For all of the instruments and mission scenarios described here, we saw an improvement in the number of total and unique detections when the minimum completeness was raised from zero.  Specific improvements varied between instrument and spacecraft designs, but all systems performed better with a minimum completeness of 0.1, as opposed to using the full target list.  In cases where low completeness stars (completeness  $< 0.1$) did provide detections, these were always very dim, required much longer integration times to acquire a spectrum, and were never detected again.  The integration time cutoff turned out to be harder to optimize due to the larger search space and the computational costs associated with running many simulations.  We chose a 50 day integration time cut-off for all missions as this value produced the best results of the ones tried; nevertheless, this clearly represents another candidate for more careful optimization in future studies.  As a final note, the inclusion of a maximum integration time carries with it an assumption that our instrument and observatory is capable of performing continuous observations for at least this length of time.  If a specific instrument design leads to a systematic limit on the length of observations, then this limit would have to be used as the maximum integration time.

\section{Results}\label{sec:results}
The purpose of this section is to demonstrate complete analyses using the techniques of \S\ref{sec:constructDRM}.  While the framework described here is fully intended to be able to incorporate all of our present knowledge about the distributions of planets in our universe, for the purposes of analyzing the instruments described below, all of which have the stated goal of finding Earth-like planets, we start with the simplified planetary population described in section \ref{sec:genPlanSys}:  All of the generated universes are populated with Earth-twins distributed throughout the habitable zones of the target systems (defined as $a \in [0.7, 1.5]$, scaled by the square root of the target star's luminosity, and $e < 0.35$).  The term `Earth-twin' here is taken to mean planets of the radius and mass of the Earth, with a constant albedo of 0.26 (this value was selected to conform with previous studies, including \citet{lindler2007}) with isotropically distributed orbital orientations.  Exo-zodi levels are taken to be constant, with $\mu = 1.55$ to correspond to the historical average zodi brightness. \citep{kuchner2008}  We also restrict our systems to at most one planet, to ensure that these missions can achieve their requirements even if Earth-like planets happen to be rare in our universe. It should be noted that the detection of any of the other planet types would be  simpler, and increasing the number of planets per system would increase the probabilities of detection.

\subsection{Validating the Framework} \label{sec:validation}
When comparing the results of mission simulations based on the decision modeling described in \S\ref{sec:decisionModeling}, it is necessary to determine whether the cost function described by equation (\ref{eq:costfunc})  actually produces optimal (or close to optimal) results.  Of course, the definition of optimality in this case depends entirely on what we want the mission to achieve, and how we prioritize the various science deliverables (i.e., how we set the weight factors $a_i$ in the cost function).  Still, if the cost function operates as designed, and all weights are set to be equal, we would expect the resulting simulations to maximize science yield while minimizing operational cost.  Since there is a direct tradeoff between time spent on new detections and time spent on characterizations, and between fuel use and available observation time, we do not expect a globally optimal solution for any of these.  Instead, we want to always be able to find the easily detectable planets in every universe, and to get orbits and spectra for as many of these as time permits, while reducing fuel use in designs with occulters.  To test whether our framework accomplishes this, we can generate one single universe, and compare the results of the mission simulation for this universe using our automated decision modeling to mission simulations with randomly generated visit orders.  Specifically, rather than using our decision modeling algorithm to decide on the next target, we simply select one at random from the list of targets currently outside of the keepout zones.  By running this randomized visit order mission on the same universe multiple times, we can map the space of possible observing sequences and find a distribution of mission results.  We can also use a modified version of our framework that always selects the next available highest completeness target, rather than using the cost function in equation (\ref{eq:costfunc}).

\begin{figure}[ht]
\plottwo{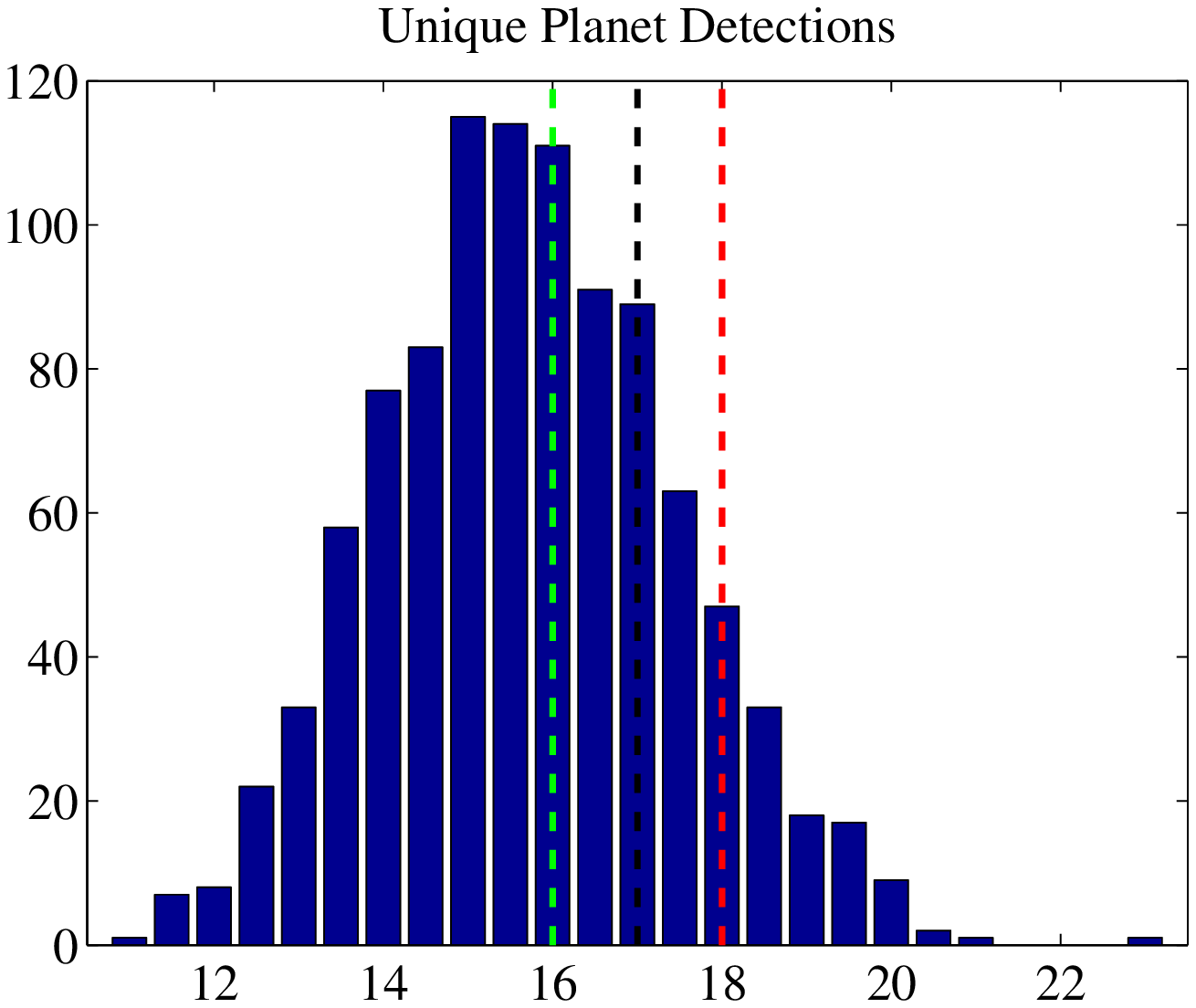}{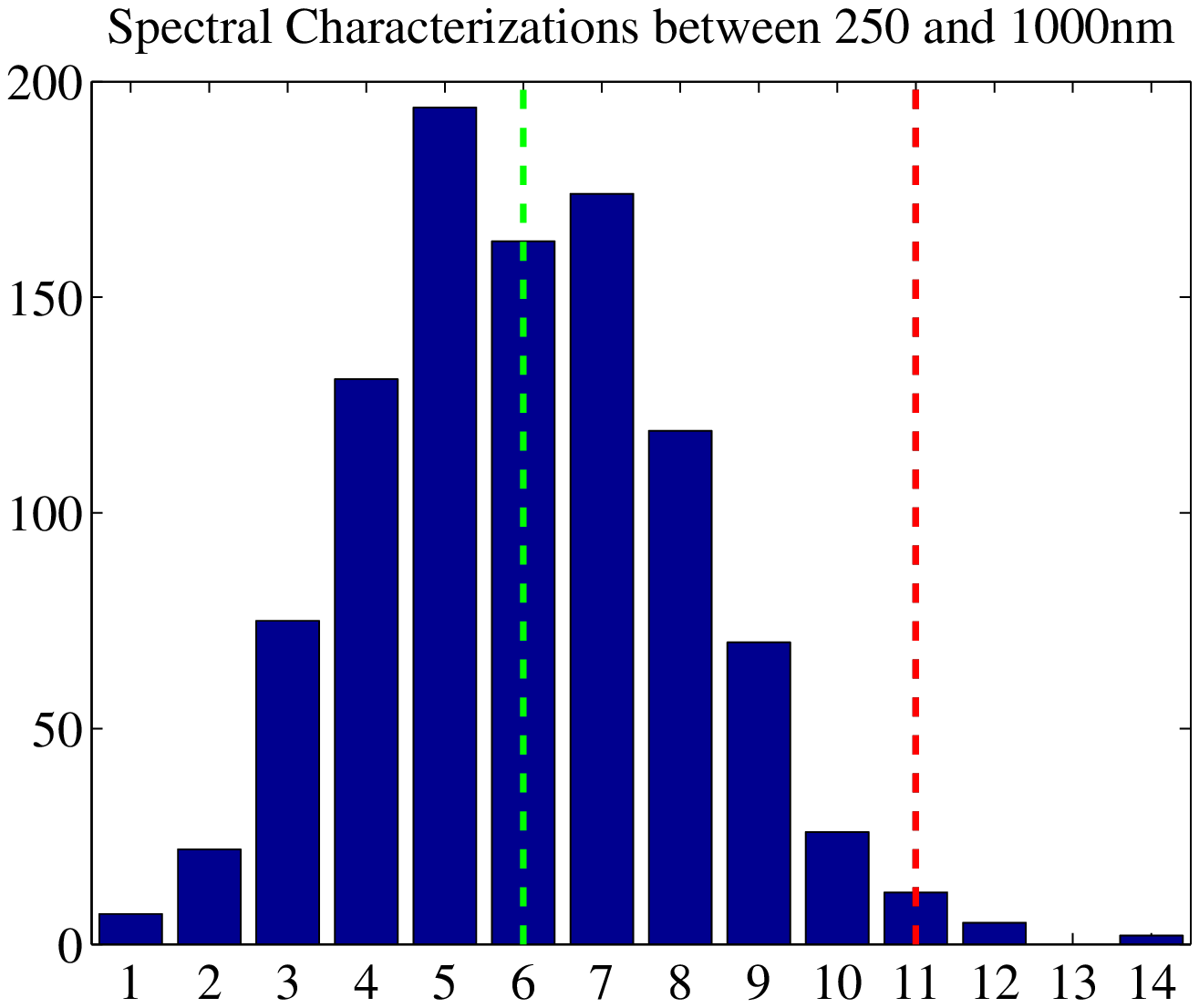}\\
\plottwo{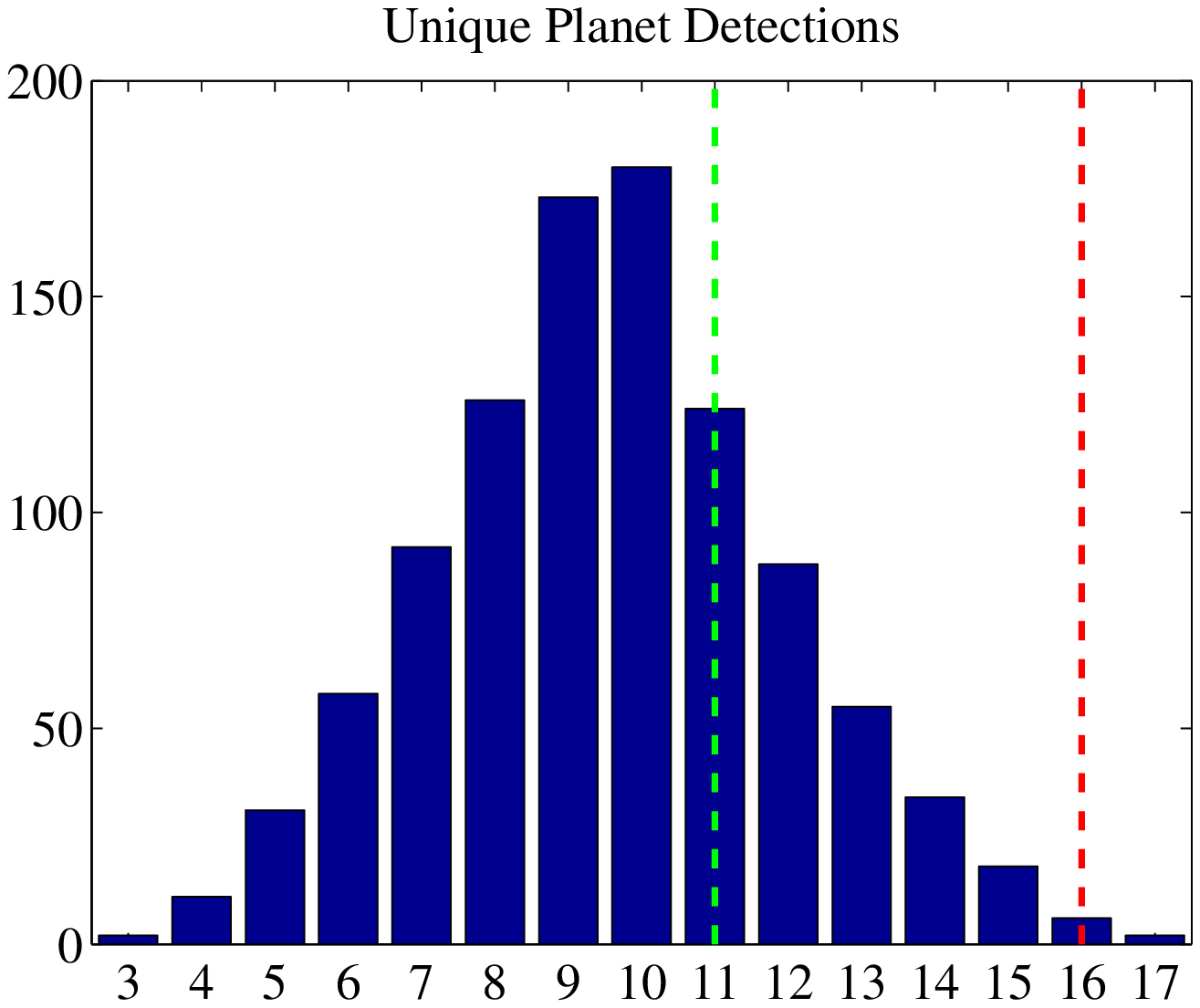}{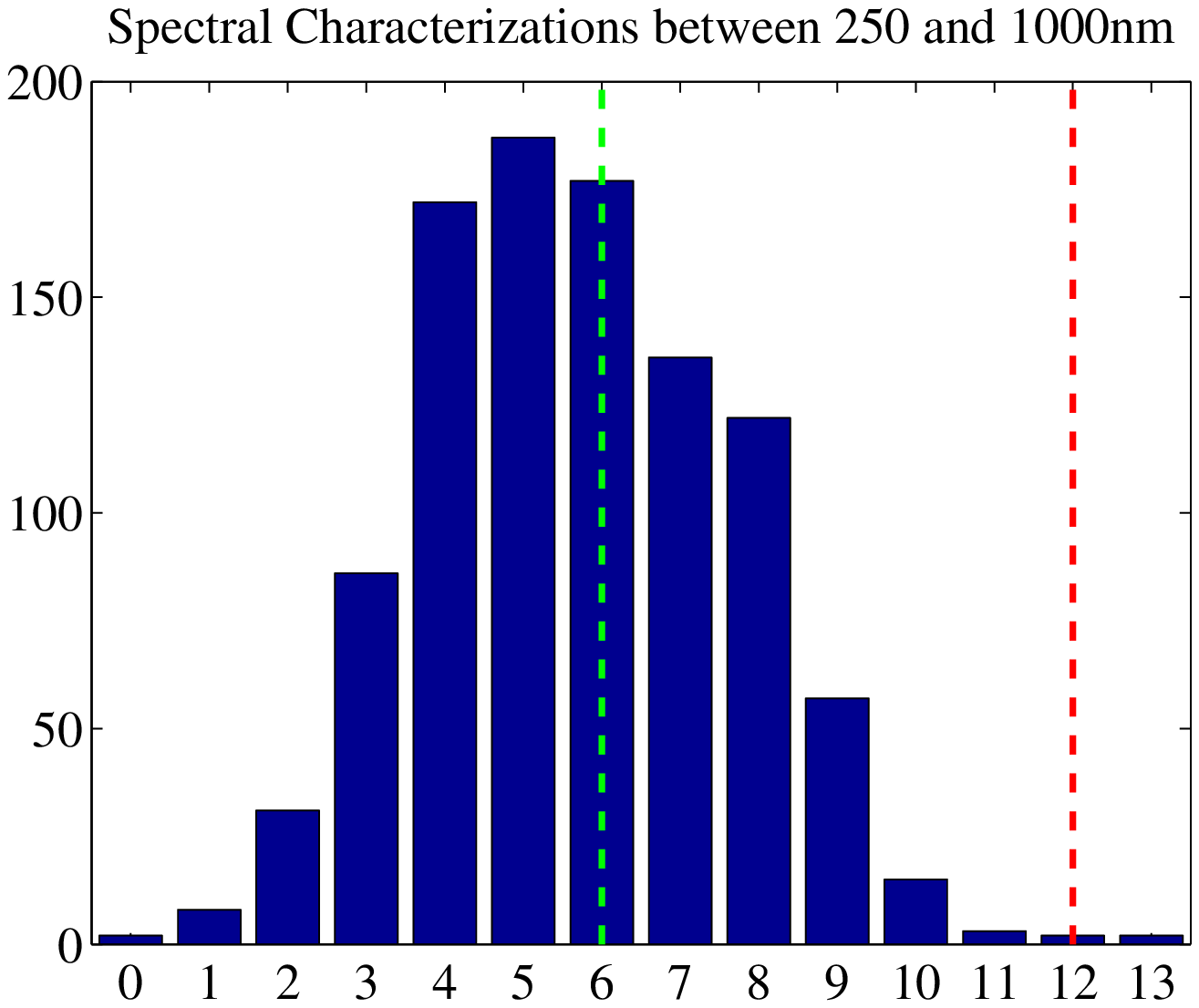}
 \caption[]{ \label{fig:optim_hists} Comparison of science yield from automated and randomized visit order selection for 2 $\lambda/D$ coronagraph (top row) and THEIA (bottom row).  Blue bars are histograms of results from 1000 mission simulations using randomized visit order. Black dashed lines are results from the automated visit order with depth of search $k = 1$, red lines are results for automated visit order with $k = 3$, and  green lines represent timelines generated by selecting the next available highest completeness target.  In the spectral characterizations for the coronagraph, and both metrics for THEIA, the $k=1$ and $k=3$ cases produced identical results.  See text for more details.}
 \end{figure}

Figure \ref{fig:optim_hists} shows the resulting histograms for the number of unique planet detections and successful spectral characterizations found from running 1000 randomized visit order missions on a single universe with $\eta_\oplus = 0.5$, using the THEIA and 2$\lambda/D$ coronagraph mission concepts (described in detail in the next section).  On top of these, we show the results of the missions produced by our decision modeling algorithm with a depth of search ($k$) of either 3 or 1, 
 and the results of the mission which only considers completeness in selecting visit order.  
For THEIA, using a $k$ of 3 and 1 produces the same values for both these metrics, whereas the $k=3$ case outperforms the $k=1$ case in terms of unique detections for the coronagraph. All cases using the scheduling algorithm described in \S\ref{sec:decisionModeling} outperform schedules based on just completeness.

\begin{figure}[ht]
 \plotone{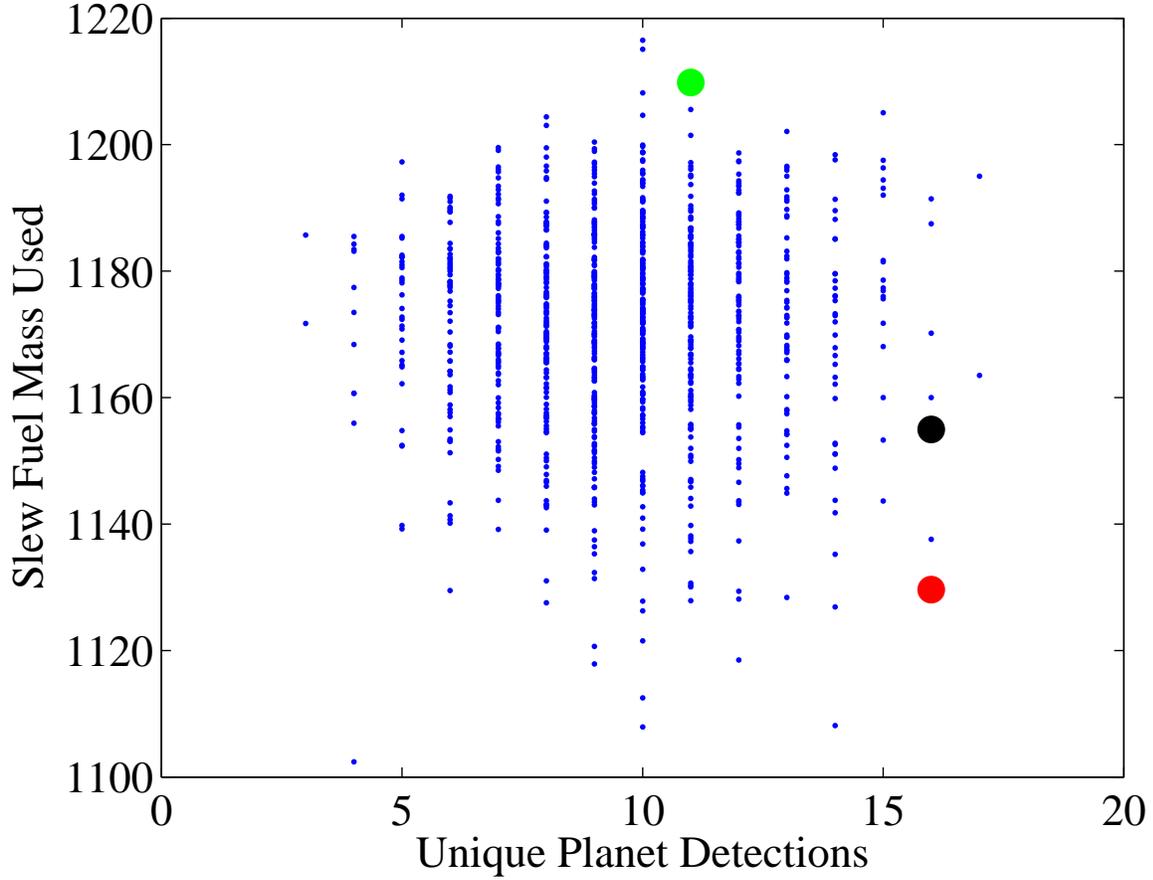}
 \caption[]{ \label{fig:optime_fuelvudets} THEIA occulter propellant use (in kg) vs. the number of unique planet detections for 1000 mission simulations using randomized visit order in one simulated universe.  The red point represents the mission timeline generated using the automated visit order for the same universe for $k = 3$, and the black point for $k = 1$.  The green point represents the timeline generated by selecting the next available highest completeness target.}
 \end{figure}

Figure \ref{fig:optime_fuelvudets} shows a scatterplot of THEIA occulter fuel use vs. the number of unique planets found for the 1000 randomized visit order missions and the automated visit order missions.  Here we do see a difference between the $k = 3$ and $k = 1$ simulations, with $k = 1$ producing a mission timeline where more fuel is used.  We compare only the amount of fuel used in slews, rather than total fuel use, since the stationkeeping fuel is directly linked to the number of observations and spectra and is automatically maximized along with these metrics.  This validation technique was applied to multiple randomly generated universes, with varying values of $\eta_\oplus$, producing similar results in each case.

These results show that the automated mission does not represent a global optimum for these metrics.  There are a small number of missions which find more unique planets, or get more spectra.  However, we can see that the automated mission does a very good job of balancing these metrics and maximizing (or minimizing, in the case of fuel use) them jointly.  Allowing for fact that the automated mission needs to be generated only once per universe, whereas a random walk search for local optima takes many trials, the benefits and efficiency of this approach become quite clear.  In terms of real mission planning, a random walk approach is simply not an option (since we only get one chance at scheduling the mission).  We also see that even when using a $k > 1$ (looking multiple steps into the future) gives us the same science yield as only considering one step, we still end up with a lower occulter fuel cost.  Finally, it is important to note that the margin between the unique planet detections using our scheduling algorithm versus the selection of the next available highest completeness target is smaller for the coronagraph than for THEIA, and is further from the tail of the distribution of randomized visit orders.  This indicates that the coronagraph performance could be further improved with optimization of the cost function weights, or by selecting a higher $k$ value.

\subsection{Comparing Instrument Designs} \label{sec:compareInstruments}
\begin{figure}[ht]
\plotone{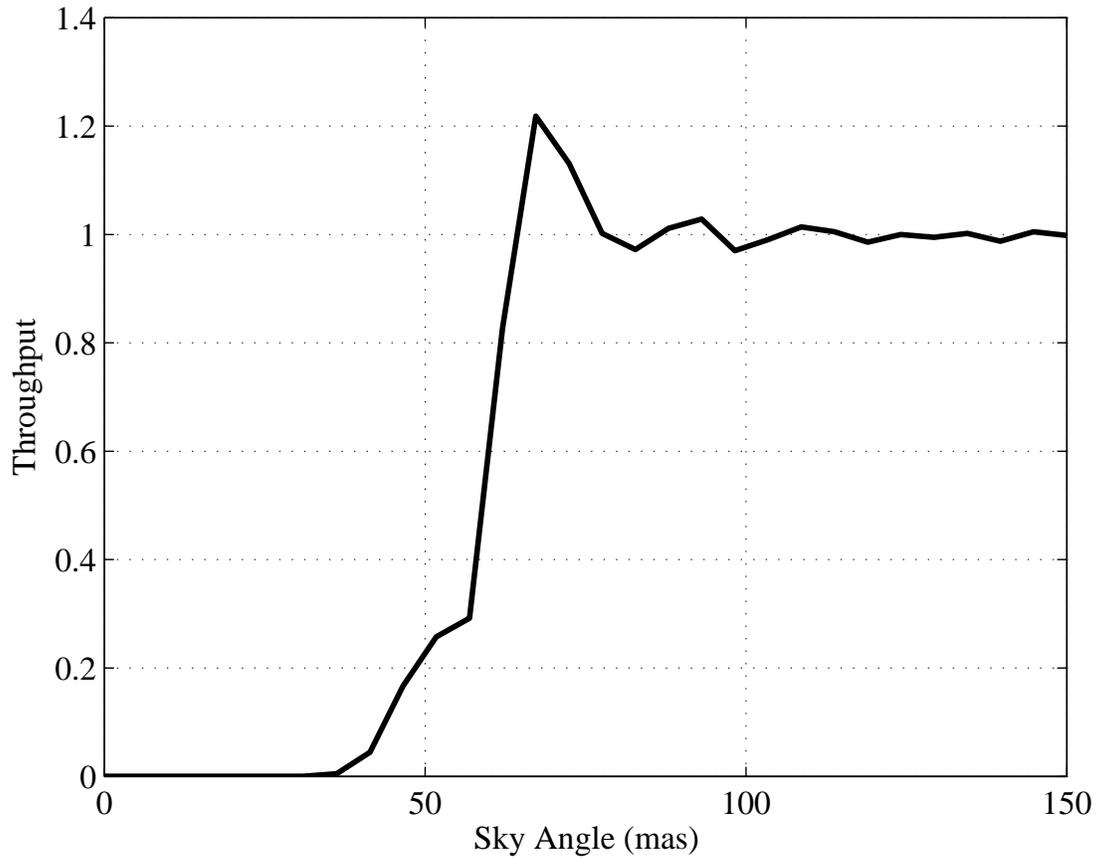}
 \caption[]{ \label{fig:sdocculter} Occulter throughput as a function of angle on the sky at 500nm.  The throughput may be greater than unity since at certain angular separations, the occulter petals will send planet light into the telescope aperture that would otherwise have bypassed it.}
 \end{figure}

\begin{figure}[ht]
  \plotone{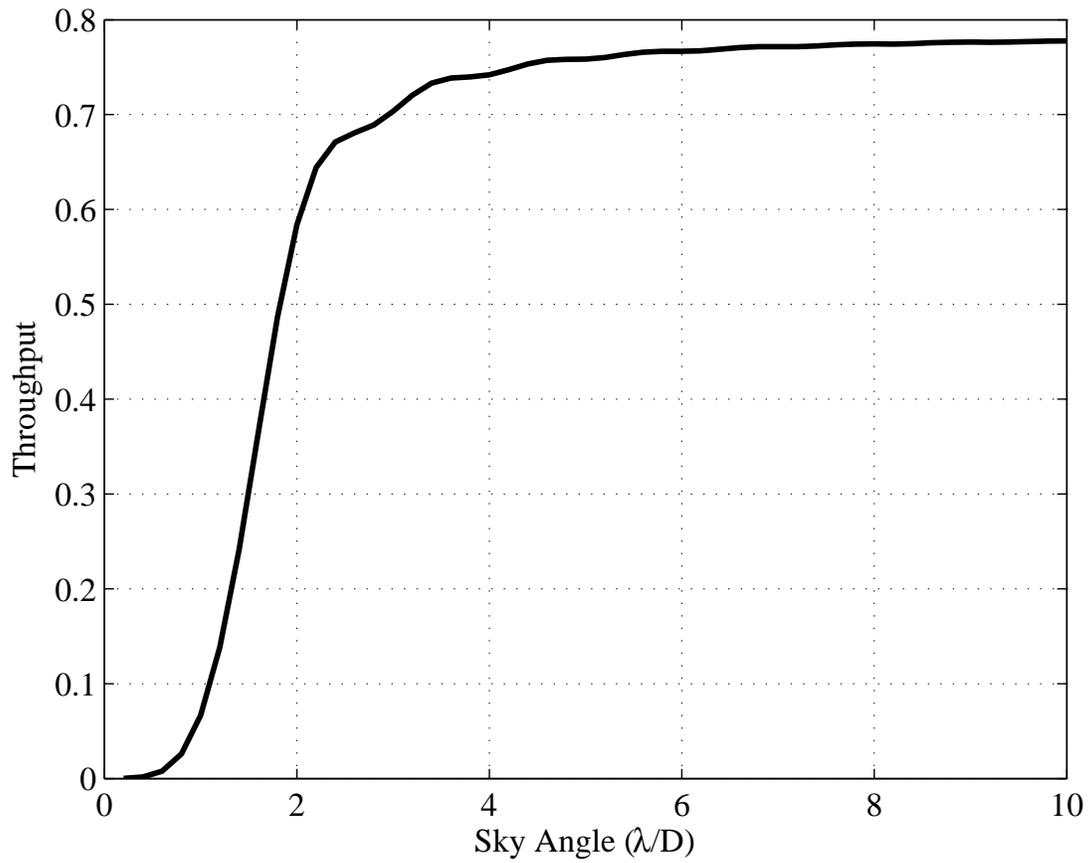}
 \caption[]{ \label{fig:coron} Coronagraph throughput as a function of angle on the sky in units of wavelength over telescope diameter ($\lambda/D$). (L. Pueyo,  Personal Communication, 2009) }
 \end{figure}

\begin{figure}[ht]
 \plotone{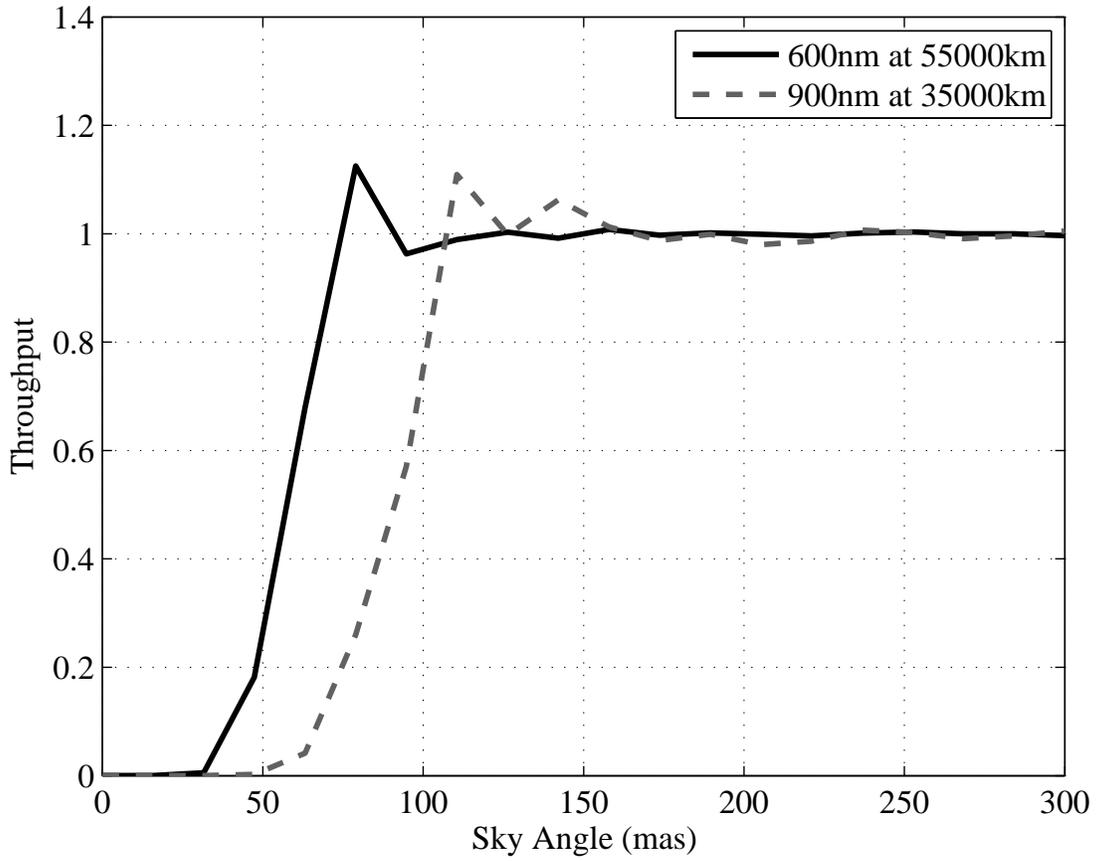}
 \caption[]{ \label{fig:theia} THEIA XPC throughput as a function of angle on the sky at 600nm with 55,000km separation and at 900nm with 35,000km separation. The throughput may be greater than unity since at certain angular separations, the occulter petals will send planet light into the telescope aperture that would otherwise have bypassed it.}
 \end{figure}

We will compare four coronagraph and occulter mission concepts designed to acquire exoplanet spectra between 250 and 1000nm, each using a telescope with a 4m diameter.  The first occulter, designed to provide high contrast across the entire spectral band, is 51.2m in diameter with a nominal separation distance of 70,400km, giving the system an angular size on the sky (geometric IWA) of 75mas.  The IWA as determined from the 50\% throughput point is 58.9mas (see Figure \ref{fig:sdocculter}). The coronagraphs are designed to have IWAs of either 2 or 3 $\lambda/D$ (wavelength/telescope aperture diameter) across the spectral band, which translates to a range between 26 and 103mas (Figure \ref{fig:coron}).  For detections in the V band this translates to an effective IWA nearly identical to that of the occulter.  The final instrument is the THEIA mission concept Exoplanet Characterizer (XPC).  While this instrument originated as an attempt to find a hybrid approach, including both a coronagraph and an occulter element, practical difficulties made this idea untenable.  The coronagraphic element was removed from the system and the final design for this instrument is a smaller (40m diameter) occulter at a separation distance of 55,000km (geometric IWA is again 75mas and 50\% throughput IWA is 57.6mas).  This comes at the cost of providing high contrast only between 250 and 700nm.  To cover the rest of the spectral band, the occulter is moved 20,000km closer to the telescope, producing a larger geometric IWA of 118mas (Figure \ref{fig:theia}).

Although moving THEIA's occulter to complete spectral characterizations introduces high costs in terms of fuel and time, there is no effective way of minimizing these, so they need not be included in the transition cost function.  Since covering the whole spectral band is a mission requirement and observing seasons for most stars are relatively short, we have no choice but to move the occulter in at full thrust if characterizing the whole spectral band is to be attempted during a single visit.  Thus, the fuel/time cost of this extra slew is essentially constant (changing only with the changing spacecraft mass) and does not affect the transition cost calculation any more than spectral characterizations performed by a coronagraph.  The only changes this system requires in the simulation logic are as follows: The occulter is moved for spectral characterization only if the star will remain observable for enough time to perform the slew and the second spectral integration.  In cases where the first half of the spectrum is acquired, but not the second, on revisits the occulter is slewed directly to the closer separation distance.  Because the smooth trajectories between to the two separation positions do not take significantly different amounts of time or fuel, the heuristic term in the original cost function still applies.

The point spread functions for the optical systems of these instruments are generated using different algorithms.  A 2 $\lambda/D$ coronagraph, such as the one modeled here, would most likely have to be a pupil mapping system.  (For details, see \citet{guyon2006})  Throughput curves for the coronagraph alone are created by sequentially generating tilted wavefronts at the telescope aperture and applying propagation algorithms to these.  For occulters, we use the Bessel expansion described in \citet{vanderbei2007} via a fast algorithm to calculate the electric field at the aperture of the telescope, assuming an incident plane wave on the occulter.  The field at the telescope aperture following a tilted incident wave may be related to the field from an incident plane wave by:
\begin{equation}
    E_{\mathrm{tilt}}(x, y) = e^{-i k \sin{\alpha} x} e^{i k z \cos{\alpha}} E(x + \sin{\alpha} z, y)
\end{equation}
which holds for small angles and allows the Bessel expansion for plane waves to be used. This field may then be fed into the appropriate propagation scheme.  (Propagation through the telescope with no coronagraph is treated here as just a Fourier transform.)  In general, we prefer to examine throughput in terms of total energy in pupil planes rather than image planes, as the finite extent of the pupil planes ensures a more accurate measure of energy. \citep{kasdin2006}  It is important to note that the throughput curves shown in Figures \ref{fig:sdocculter} - \ref{fig:theia} are idealized, and achieving these levels of performance will require meeting exacting manufacturing and positioning tolerances for the occulters, and very precise wavefront control for the coronagraphs.  These considerations are beyond the scope of the current study, but must be considered in the future.

For each design, 100 simulations of a complete 5 year mission were generated at equally spaced values of $\eta_\oplus$ between 0.1 and 1. Table \ref{table:compinsts} lists the values of the other simulation parameters.  Mission rules were kept fixed between designs, where applicable.  For all four instruments, mission rules were set to prioritize revisits to systems with prior detections up to 4 total detections of the same planet.  Only one full spectrum was required for each unique planet, so no time was spent on characterization during revisits if a spectrum had been acquired in the initial detection.  For the two designs featuring occulters, target completeness was prioritized at half the value of transit costs ($a_1 = 2a_2$), and transit slews were only allowed in cases where at least half of the slew time would be spent thrusting. The requirement for successful spectra was set to a S/N value of 11 for a spectral resolution of 70.  In addition to integration and occulter slew times (where appropriate), 24 hours of spacecraft overhead time were added for each target system visited.  The coronagraph designs were allowed to use 50\% of the full mission time for planet-finding operations, whereas the occulter designs used all available time not spent on transit slews (approximately 1/3 of the mission time, on average).

\begin{figure}[ht]
   \plottwo{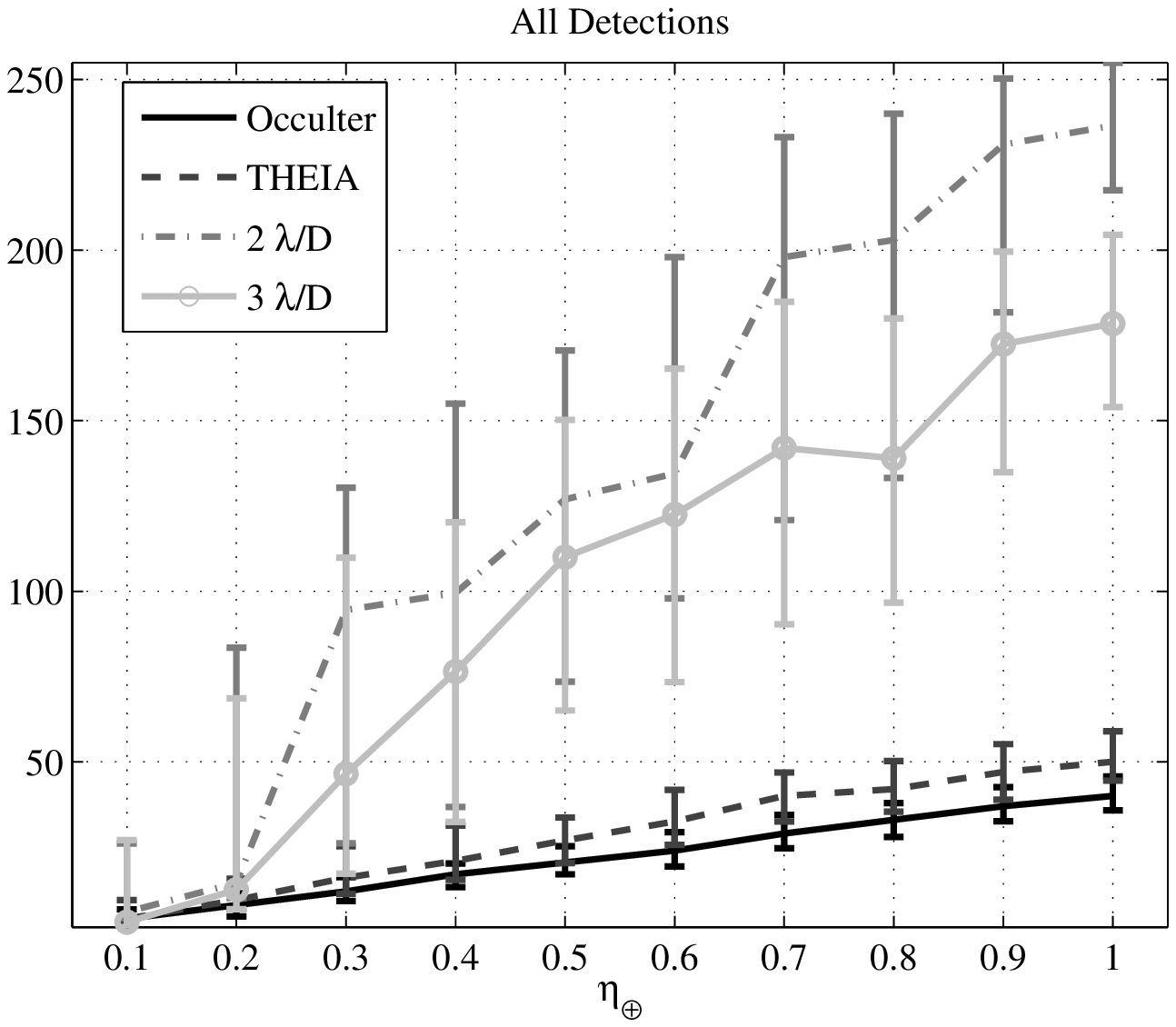}{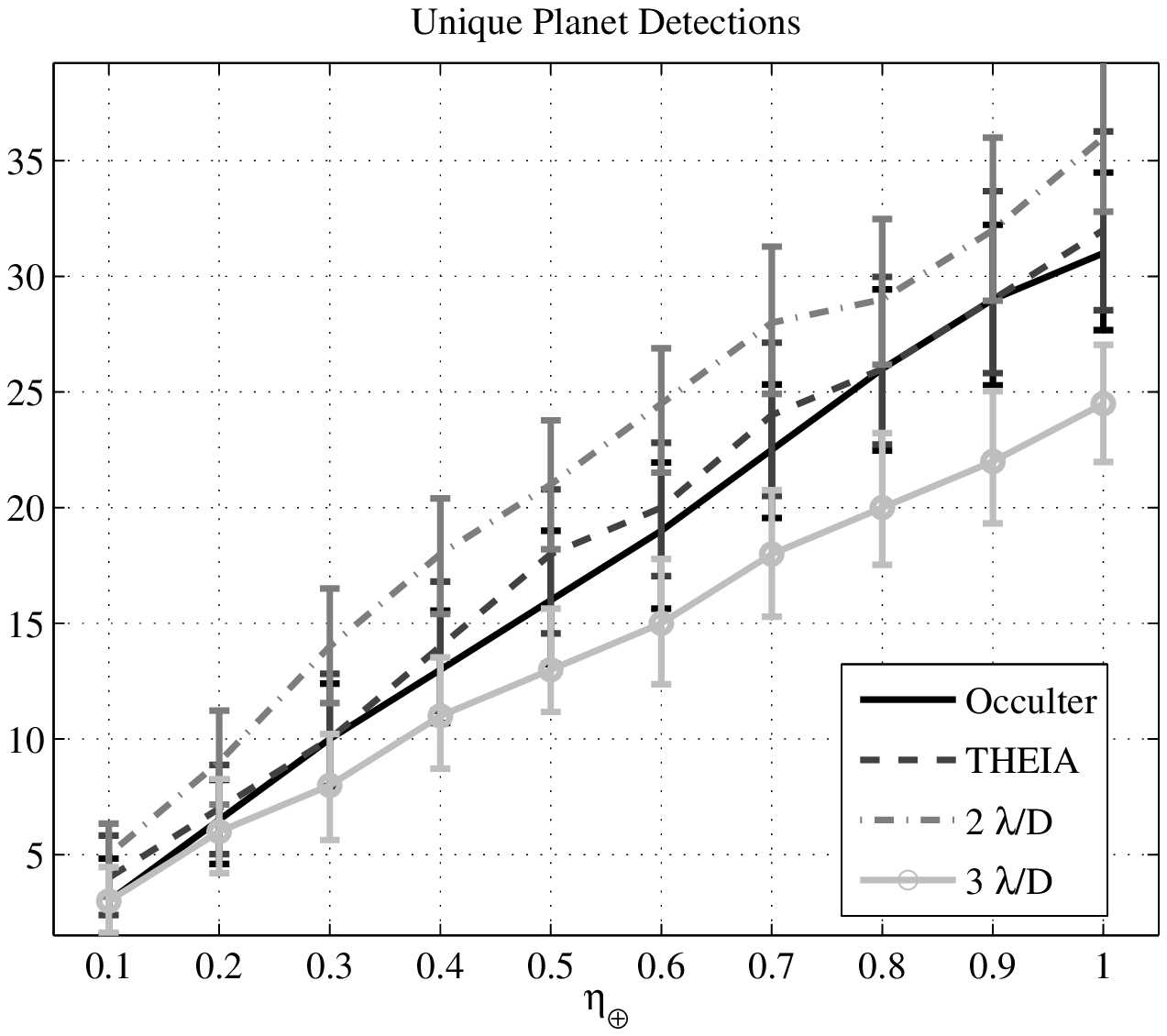}\\
   \plottwo{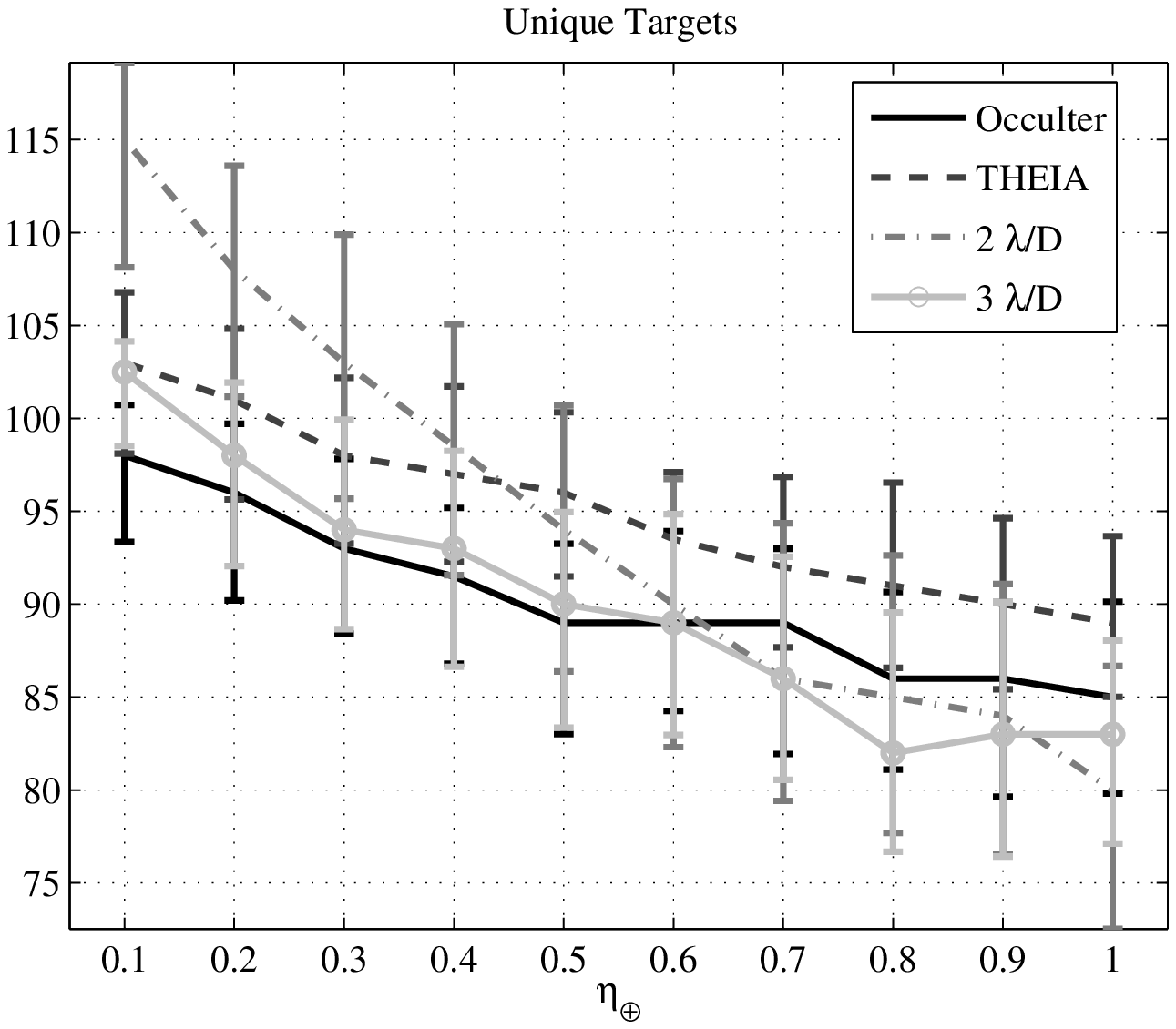}{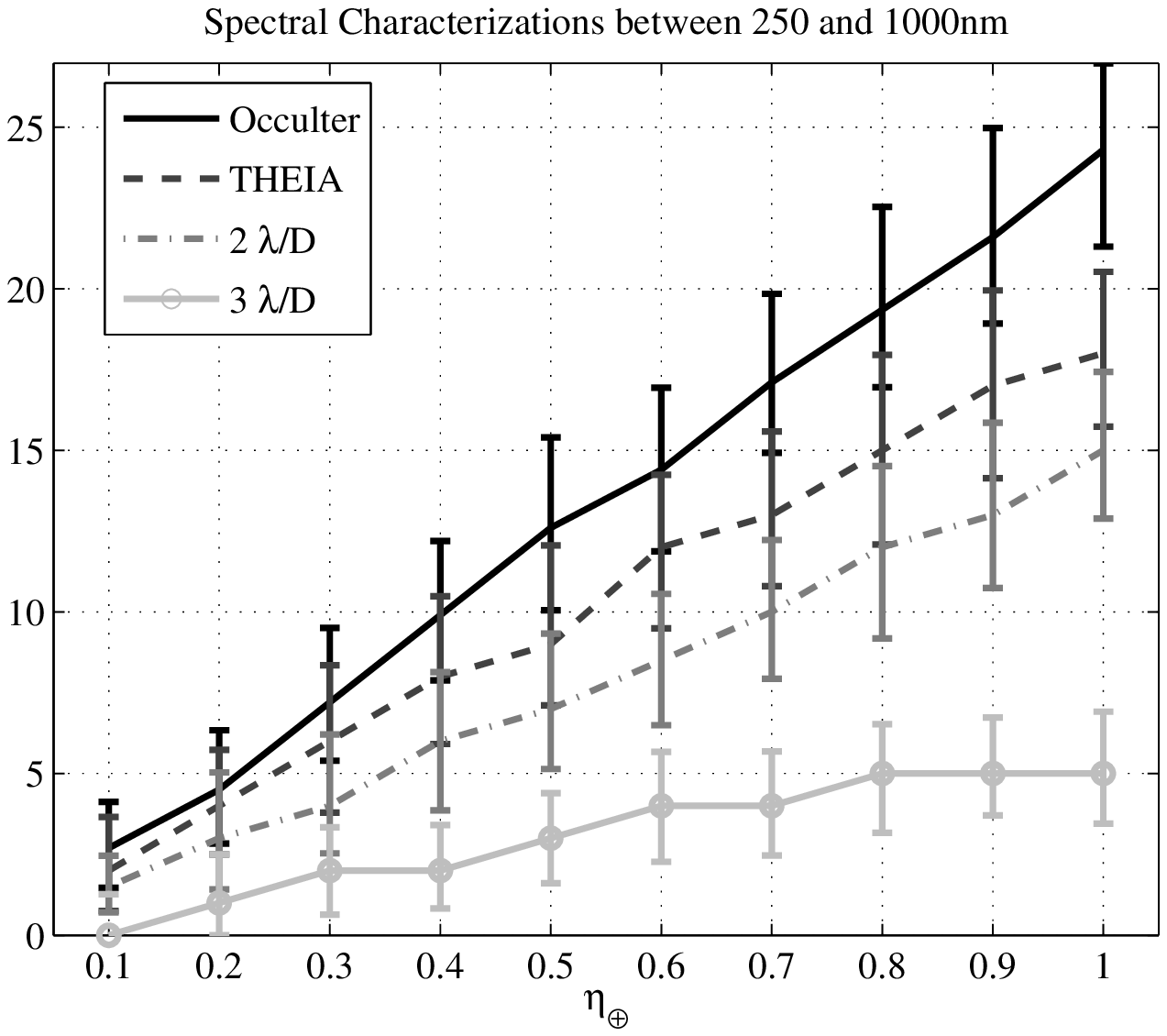}
 \caption[]{ \label{fig:compinsts1} Simulation results for occulter, THEIA, and the 2 and 3 $\lambda/D$ coronagraph mission concepts as functions of $\eta_\oplus$. The lines represent median values for 100 simulations at each value of $\eta_\oplus$.  Errorbars represent the one-sided deviations of each distribution.  The top left plot shows the total number of planets found (including multiple detections of the same planet), the top right plot shows the number of unique planets found, the bottom left plot shows the number of unique target systems visited during the mission, and the bottom right plot shows the total number of complete spectra (250-1000nm) acquired.}
 \end{figure}

Figure \ref{fig:compinsts1} shows the science results based on the metrics of: total number of planetary detections, number of unique planets detected, total number of unique target stars observed, and the number of full (250-1000nm) spectra acquired.  The results shown here are the median values from 100 full mission simulations at each value of $\eta_\oplus$ (between 0.1 and 1 in increments of 0.1).  The plotted errors correspond to the one sided deviations for the distribution halves above and below the median values.  The most striking feature of these results is the strong linear dependence of all on $\eta_\oplus$.  We note, however, that the slopes of the various lines vary significantly between different instruments.  It is therefore incorrect to infer results for an arbitrary $\eta_\oplus$ from results at only one other $\eta_\oplus$, unless an analysis such as this one has been previously completed.  In order to extrapolate results for a range of $\eta_\oplus$ values, it is necessary to simulate missions for at least two $\eta_\oplus$ values.  While it has been suggested that only one value need be evaluated, since all of these metrics must (by definition) equal zero when $\eta_\oplus$ is zero, least squares linear fits to some of the data sets in Figure \ref{fig:compinsts1} have significant non-zero intercepts.  It is possible that the linear dependence of these metrics on $\eta_\oplus$ breaks down (or changes slope) in very low $\eta_\oplus$ cases.  Alternatively, it may be that the dependence is not strictly linear, but has a strong linear component in the range of $\eta_\oplus$ values being considered here.  There is no rigorous demonstration that all of these metrics should be strictly linear in $\eta_\oplus$ so this certainly represents an area in which more work is required.

The other important feature of these results is that they reveal that for a 4m telescope, the occulter and coronagraph mission concepts are highly competitive, with specific strengths and weaknesses.  In terms of total number of detections, the coronagraphs are the clear winners, with many times as many detections as either of occulter designs.  These results may be slightly misleading, since they represent, for the most part, many repeated detections of a small number (usually fewer than 5) highly visible planets.  Still, this clearly demonstrates the advantage of being able to perform many more observations over the course of the mission and, with improvements to the scheduling algorithm, should be translatable into a larger number of orbital fits for the coronagraphs (which is a function of the number of repeated observations, as long as the planet is observed at different points on its orbit each time).  The two occulters average less than two detections per unique planet found, which means that they are generally unable to get more than one orbital fit in the five years simulated here.

\begin{figure}[ht]
\plotone{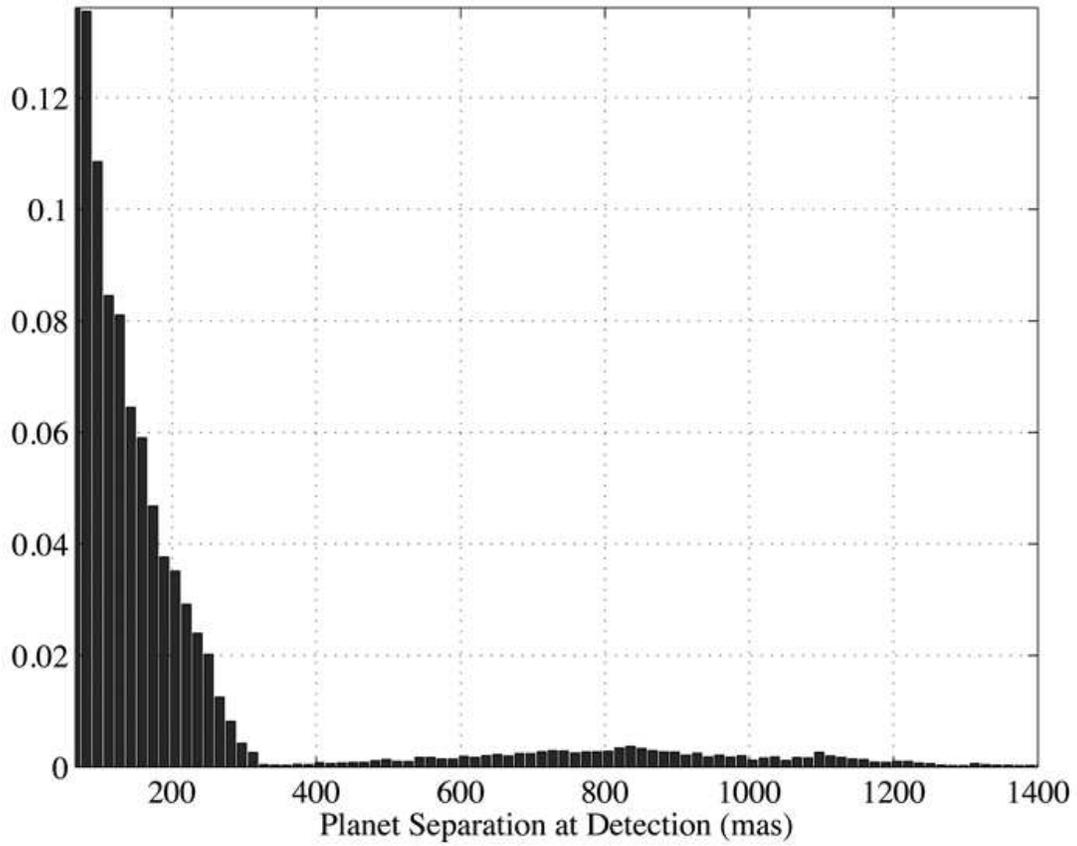}
 \caption[]{ \label{fig:detIWAs} Histogram of planet separation angles at time of detection for 37000 successful detections by the coronagraph design in the simulations shown in figure \ref{fig:compinsts1}. The shape of the histogram is due to the assumed distribution of planets and orbits.}
 \end{figure}

In terms of unique planet detections, the results are much more even, with the occulters performing almost identically, and finding about 10\% fewer planets than the 2 $\lambda/D$ coronagraph and 25\% more than the 3 $\lambda/D$ coronagraph at $\eta_\oplus = 1$.  This is almost purely a function of having a limited target list.   In a universe with a uniform distribution of isotropically positioned stars, we would expect the coronagraphs' larger total number of observations over the mission lifetime to give them a significant advantage.  However, using only the limited target list available, all of the instruments are able to find all of the easily observable planets within five years.  
All of the mission concepts consistently find fewer than half of the available planets in a given universe.  The coronagraphs' large number of detections is balanced by their relatively poor performance in terms of spectral characterization.  Because a coronagraph's IWA is wavelength dependent, and the majority of detections for Earth-like planets occur at relatively low separations (Figure \ref{fig:detIWAs}), in most cases the coronagraphs are capable of only a partial spectrum between 250 and 1000nm.  We see that the $2\lambda/D$ design gets slightly fewer full spectra than THEIA (which has difficulties getting full spectra due to the extra required slew for the longer wavelength characterizations), but both are outperformed by the single distance occulter.  The 3 $\lambda/D$ coronagraph is highly IWA limited, and gets many fewer full spectra than any of the other designs.  From these metrics, we can conclude that for a 4m telescope, coronagraphs and occulters are both viable options, and the decision as to which instrument to build should be made based on other considerations, including the relative difficulties of wavefront control versus starshade manufacturing and positioning.  It is also important to keep in mind that the transit slews required by the occulter designs automatically make more than half of the mission time (2/3 of the mission on average, depending on the scenario), available for general astrophysics and other science.  In these simulations, the coronagraph designs were allowed 50\% of the mission time for planet-finding, consistently giving them more total planet science related integration time.  Equal time could be allocated for the coronagraph designs, at the cost of some planetary detections.
 
\begin{figure}[ht]
\plotone{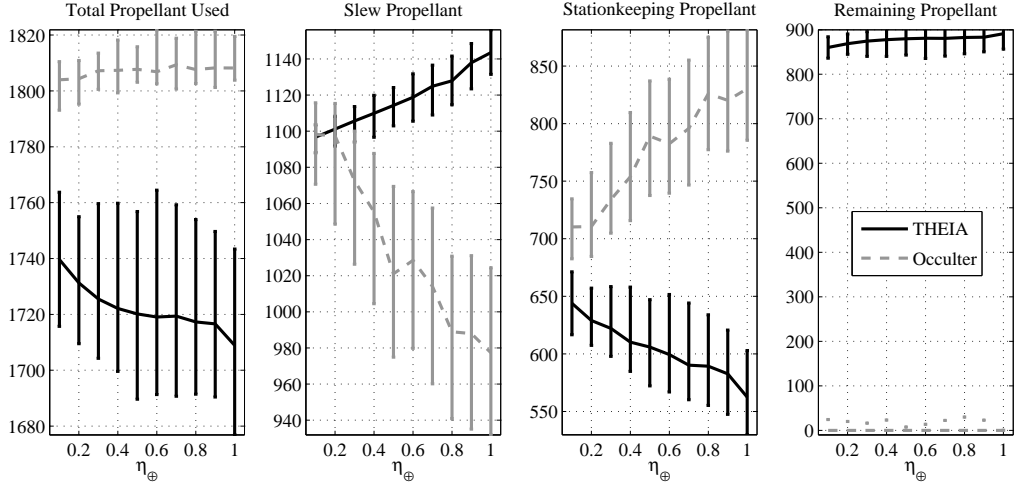}
 \caption[]{ \label{fig:fuel_use} Occulter propellant use (in kg) for designs including an occulter in the simulations shown in figure \ref{fig:compinsts1}. Although the assumed stationkeeping and slewing propulsion systems use different propellants, the simulation does not distinguish between the two fuel types and simply assumes that enough fuel is available, up to a maximum total fuel mass.  When this mass is exceeded, the mission ends, regardless of mission time. The primary cause for the difference in the fuel remaining after 5 years is the difference in starshade dry masses between the two designs.}
 \end{figure}

 In addition to science yields, the simulation can also be used to evaluate engineering requirements.  One of the most important of these is the capability of the spacecraft propulsion systems to support the instrument science operations.  This is particularly important in the case of designs including occulters, since the occulter must carry enough fuel to move from target to target, and also to maintain a highly precise position during integration.  Figure  \ref{fig:fuel_use} shows the average propellant use for the simulations of the occulter and THEIA mission concepts.  Both occulters were assumed to have the same initial mass (the payload capacity of the target launch vehicle), which translates to a smaller total fuel capacity for the heavier, single distance occulter.  Slewing was simulated as performed using high-efficiency electric propulsion ($I_\textrm{sp}$ of 4160 s) while stationkeeping used a chemical propulsion system ($I_\textrm{sp}$ of 220 s).\footnote{A separate system was added for stationkeeping due to concerns that the plumes produced by the electric propulsion system would be bright enough to interfere with observations.}   As expected, the single distance occulter, having the larger separation distance from the telescope, uses more fuel overall.  It is very interesting, however, to look at the breakdown of fuel use in terms of slewing and stationkeeping.  Because characterizations for THEIA include an extra slew at full thrust, as $\eta_\oplus$ increases so does the number of characterizations and the slew propellant usage.  In the case of the single distance occulter, on the other hand, more time characterizing translates to less time slewing, and so its slew propellant use drops with increasing $\eta_\oplus$.  A similar effect is seen in the stationkeeping propellant use - as $\eta_\oplus$ increases, THEIA spends more and more time at the smaller separation distance, where stationkeeping fuel costs are lower, thereby slightly lowering overall stationkeeping fuel use.  At the same time, the single distance occulter spends more time characterizing at its wider separation, increasing its stationkeeping fuel use.  All of these effects, however, are relatively minor when compared with the difference in the two occulters' dry masses---the fuel remaining at the end of THEIA's primary mission is due almost completely to the fact that the occulter is smaller and lighter than its single distance counterpart.
 
It now makes sense to see if we can take advantage of the extra fuel at THEIA's disposal in an extended mission to make up for some of the differences in science yield between the two occulter designs.  One option is to continue into the extended mission with the same mission rules as for the primary mission.  However, because there is generally only about 2.5 years of fuel left, and the THEIA design is able to find all of the easily observable planets during the primary mission, this approach does not significantly increase overall science yield.  Another alternative is to limit the target list in the extended mission to only those stars which had planets detected in the primary mission.  This allows for many additional detections, and, in some cases, a few more complete spectra.  

\begin{figure}[ht]
\plotone{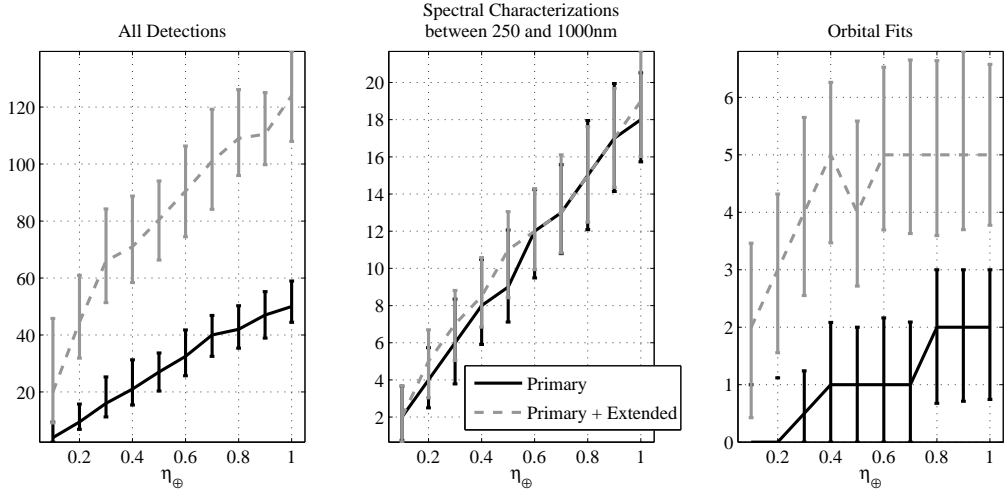}
 \caption[]{ \label{fig:theia_ext} Simulation results for the primary THEIA mission shown in figure \ref{fig:compinsts1}, and the primary + extended mission, where the occulter is allowed to use all of the available fuel.}
 \end{figure}

Figure \ref{fig:theia_ext} shows the results for the primary and primary + extended missions.  Unfortunately, for this design, the extended mission does little to increase the number of complete spectra attained, mostly because the characterization times allow for full spectra only for planets that are observable for longer than the average observing season for this population.  This means that if a full spectrum was achievable, it would generally have been acquired in the primary mission.  However, the limited target pool with guaranteed observable planets allows us to significantly increase the total number of detections in a relatively short time.  Enough of these occur at varying points on the planets' orbits that this translates to a significant number of orbital fits for the total mission (up to 5, on average, for $\eta_\oplus = 1$).  In this respect, THEIA is superior to the single distance occulter design.  With the larger occulter, to get the same number of orbital fits, it would be  necessary either to increase the fuel carried (thereby requiring a larger launch vehicle and increasing costs), or to partition the primary mission in this fashion, allocating the last one or two years to only revisiting systems with previous detections, which would sacrifice the total number of unique planets found.

\section{Conclusions and Future Work}
In this paper, we have demonstrated a complete framework for the simulation of direct detection planet finding missions and have used it to evaluate the performance of four potential mission concepts based on a 4m diameter telescope.  In doing so, we showed not only the large number of science and engineering metrics that can be evaluated with this framework, but also the optimality of our automated mission planning algorithm.  This demonstration also included a method for mapping the space of possible mission timelines on a fixed universe via large numbers of randomly scheduled mission simulations, which we expect will prove very useful in the future when testing new mission rule sets and scheduling algorithms.

From our simulation results, we find that at the 4m scale, the internal coronagraph and external occulter starlight suppression systems produce comparable results in terms of the number of unique planets found.  The coronagraph based instruments have a clear advantage in terms of faster retargeting times, and are thus able to produce many more total detections over the course of the mission.  On the other hand, occulters, whose inner working angle is not wavelength dependent and which have higher throughput, are able to collect more spectra over the full wavelength range between 250 and 1000nm.  We find that for this telescope aperture size, coronagraphs need to achieve inner working angles of 2 $\lambda/D$ or lower to have comparable spectral characterization results with occulter systems.  We also explore the explicit trade-off between occulter size and available mission time in the case of a fixed launch vehicle, demonstrating the benefits of a smaller occulter which can function beyond its primary mission length without the need for refueling. 

Although we are confident in the results we have presented, much work remains to be done to improve the fidelity of many of the quantities we calculate.  In each of the preceding sections, we have attempted to fully describe and justify the various approximations that were made in order to make the problem tractable without losing accuracy, but we fully expect to be able to improve on many of these in the future.  A major component of the simulation is the calculation of integration times for detection and characterization.  Currently these procedures include assumptions that exozodiacal dust is approximately  uniformly distributed, and that we can approximate the number of photons reflected from a planet using only a zero magnitude calibrated star spectrum.  Both of these assumptions should be tested---the first by assuming non-isotropic (`clumped') exozodiacal dust, and the second by incorporating model spectra for each of the target stars.  As our knowledge of exoplanets grows, and planet formation and migration theories are further developed, we can continuously improve our planet generation algorithms, considering more complex (and realistic) planetary populations.  Furthermore, by modeling the actual data we expect to collect, and the techniques to extract planet signals from this, we can more rigorously define what counts as a missed detection, false alarm, or detection, rather than simply relying on the statistical approach described here.

There remain many improvements and optimizations that can be applied to the decision modeling as well, including a rigorous analysis of the parameter space defined by the weights of our cost function.  As we gain more experience with this framework, we fully expect the planning tools used to become more and more applicable to the planning of actual missions, continuing the recent trend of increasingly autonomous spacecraft. Finally, there will always be more cases to run.  As more designs are finalized, or previous designs perfected, we can continue to test them against each other, identifying strengths and weaknesses.  For any given design there will be the challenge of finding an optimal rule set, or any rule set that achieves all of the mission's requirements.  Much also remains to be done in terms of analyzing the interaction between different mission types.  The framework described here is a perfect tool for evaluating the effects of incorporating precursor data in direct detection mission scheduling, which we are currently investigating.  Furthermore, while we have focused on a highly simplified universe populated with only Earth-twins, the most interesting results will come from analyses involving mixtures of planets and multi-planet systems.

\acknowledgments{ The authors would like to thank Don Lindler for his help with the integration time modeling and general input on constructing missions, Laurent Pueyo for providing us with accurate coronagraph models, Doug Lisman for all his work on the THEIA propulsion design, Robert Brown and Robert Vanderbei for much helpful discussion, and the entire THEIA team for their input and support.}

\bibliographystyle{plainnat}
\bibliography{Main} 

\begin{table}[ht]
\caption{Symbols and abbreviations. \label{table:syms}}
\begin{tabular}{ c  l }
\tableline\tableline
Symbol & Definition\\
\tableline
$a$ & semi-major axis \\
$e$ & eccentricity \\
$p$ & albedo \\
$\eta$ & Expected frequency of planets in the universe\\
$M_{est}$ & Estimated star mass\\
$M_{true}$ & True star mass\\
$V_s$ & Visual magnitude of the star\\
$\mathbf{r}_i$ & Position of planet $i$ with respect to the system barycenter\\
$\mathbf{r}_{i/s}$ & Position of planet $i$ with respect to the star\\
$\Delta\textrm{mag}$ & Difference in brightness between star and planet\\
IWA & Inner working angle\\
PSF & Point spread function\\
$z_\textrm{orb}$ & Telescope Orbit diameter in azimuthal direction\\
$A$ & Telescope aperture area\\
QE & Detector quantum efficiency\\
$\upsilon$ & Attenuation due to optics other than the coronagraph/occulter\\
$\Delta\alpha$ &Pixel area on the sky\\
DR & Detector dark current rate\\
$\sigma_r$ & Detector read noise\\
$\mu$ & Exo-zodi level\\
$k$ & Scheduling depth of search\\
\tableline
\end{tabular}
\end{table}

\begin{table}[ht]
\caption{Simulation Parameters. \label{table:compinsts}}
\begin{tabular}{ c  l }
\tableline\tableline
Parameter & Value or range\\
\tableline
$a$ & [0.7,1.5] $\sqrt{L}$AU\\
$e$ & [0,0.35] \\
$p$ & 0.26 \\
$z_\textrm{orb}$ & 510000 km \\
QE & 0.91(for detection)\\
$A$ & $4\pi$\\
$\upsilon$ & 0.57 \\
DR & 0.001 c/s\\
$\sigma_r$ & 3 e\\
$\mu$ & 1.55\\
\tableline
\end{tabular}
\end{table}
 
\end{document}